\documentclass{aa}  
\usepackage[multidot]{grffile}
\usepackage{graphicx}
\usepackage{float}
\usepackage{txfonts}
\usepackage[colorlinks,allcolors=blue,bookmarks=false,hypertexnames=true]{hyperref} 
\begin{document} 

   \title{Probing the filamentary nature of star formation in the California giant molecular cloud\thanks{The 18.2\arcsec\ column density map (cf. Fig.~\ref{California18p2}), and the filament skeleton map (cf. Fig.~\ref{network}) in FITS format are publicly available at the CDS via anonymous ftp to cdsarc.u-strasbg.fr (130.79.128.5) or via http://cdsweb.u-strasbg.fr/cgi-bin/qcat?J/A+A/.}}
   \author{Guo-Yin~Zhang\inst{1,2}
           \and
          Philippe~Andr\'e\inst{2}
          \and
          Alexander~Men'shchikov\inst{2}
          \and
          Jin-Zeng~Li\inst{1}
}
   \institute{National Astronomical Observatories, Chinese Academy of Sciences, Beijing 100101, PR China\\
              \email{zgyin@nao.cas.cn} 
         \and
            Universit{\'e} Paris-Saclay, Universit{\'e} Paris Cit{\'e}, CEA, CNRS, AIM, 91191 Gif-sur-Yvette, France\\
             \email{philippe.andre@cea.fr; alexander.menshchikov@cea.fr}
             }

   \date{Received; accepted}

 
  \abstract
   {Recent studies suggest that filamentary structures are representative of the initial conditions of star formation in molecular clouds and support a filament paradigm for star formation, potentially accounting for the origin of the stellar initial mass function (IMF). The detailed, local physical properties of molecular filaments remain poorly characterized, however.} 
 {Using {\it Herschel} imaging observations of the California giant molecular cloud, we aim to further investigate the filament paradigm for low- to intermediate-mass star formation and to better understand the exact role of filaments in the origin of stellar masses.}
 {Using the multiscale, multiwavelength extraction method \textsl{getsf}, we identify starless cores, protostars, and filaments in the {\it Herschel} data set and separate these components from the background cloud contribution to determine accurate core and filament properties.}
   {We find that filamentary structures contribute approximately 20\% of the overall mass of the California cloud, while compact sources such as dense cores contribute a mere 2\% of the total mass. 
   Considering only dense gas (defined as gas with $A_{V,\text{bg}} > $~4.5--7), filaments and cores contribute $\sim$66--73\% and 10--14\% of the dense gas mass, respectively.
The transverse half-power diameter measured for California molecular filaments has a median undeconvolved value of 0.18\,pc, consistent within a factor of 2 with the typical $\sim $0.1\,pc width of 
nearby filaments from the {\it Herschel} Gould Belt survey.
A vast majority of identified prestellar cores ($\sim $82--90\%) are located within $\sim $0.1\,pc of the spines of supercritical filamentary structures. Both the prestellar core mass function (CMF) and the distribution of filament masses per unit length or filament line mass function (FLMF) are consistent with power-law distributions at the high-mass end, $\Delta N/\Delta {\rm log}M\propto M^{-1.4 \pm 0.2}$
at $M > 1\,M_\odot$ for the CMF and $\Delta N/\Delta {\rm log} {M}_{\rm line} \propto {M}_{\rm line}^{-1.5\pm0.2}$ 
for the FLMF at $M_{\rm line} > 10\,M_\odot {\rm pc^{-1}}$, which are both 
consistent with the Salpeter power-law IMF. 
Based on these results, we propose a revised model for the origin of the CMF in filaments, whereby the global prestellar CMF in a molecular cloud arises from the integration of the CMFs generated by individual thermally supercritical filaments within the cloud.}
  {Our findings support the existence a tight connection between the FLMF and the CMF/IMF and suggests that filamentary structures represent a critical evolutionary step in establishing a Salpeter-like mass function.}
   

   \maketitle
%


\section{Introduction}

\begin{figure*}
   \centering
   \includegraphics[width=1.0 \hsize]{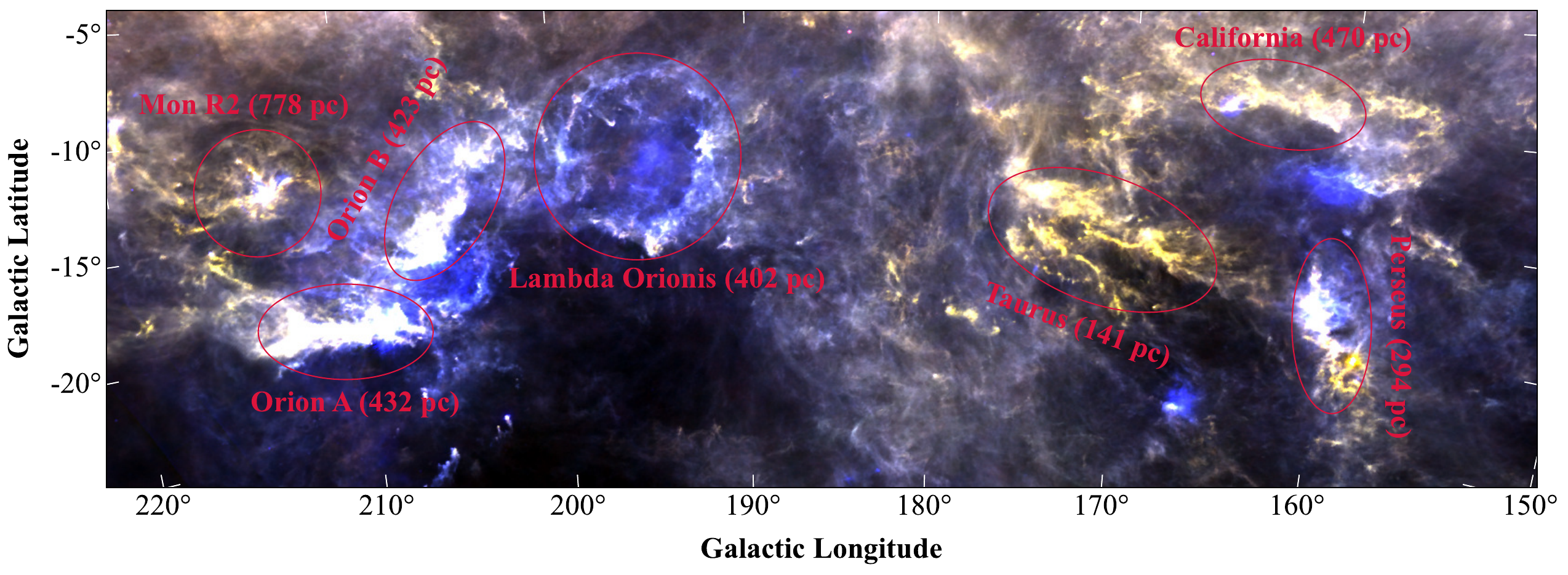}
   \caption{Location of the California GMC with respect to other nearby molecular cloud complexes. 
The three-color RGB map was created using publicly available data: 850~$\mu $m (\emph{red}) and 350~$\mu $m (\emph{green}) from {\it Planck} imaging, and 100~$\mu $m (\emph{blue}) from IRAS \citep[][Public Data Release 2]{Planck+2014}. The distances of the molecular clouds indicated on the map were inferred from {\it Gaia} DR2 parallaxes \citep{Zucker+2019}.}
\label{largemap}
\end{figure*}

Imaging surveys of Galactic molecular clouds with the {\it Herschel} Space Observatory have revolutionized our understanding of the link between the structure of the cold interstellar medium (ISM) and the star formation process, emphasizing the role of filamentary structures \citep[e.g.,][]{Andre+2010,Molinari+2010, Henning+2010, Hill+2011,Juvela+2012}. 
While the existence of filamentary structures in molecular clouds has been known for a long time \citep[e.g.,][]{Schneider+1979,Bally+1987,Myers2009},  
{\it Herschel} observations have demonstrated that filaments are truly ubiquitous in the cold ISM \citep[e.g.,][]{Menshchikov+2010} and that they make up a dominant fraction of the dense gas in molecular clouds \citep{Schisano+2014,Konyves+2015}. The observed filaments display a high degree of universality in their properties \citep{Arzoumanian+2011} and are the prime birthplaces of prestellar cores \citep[e.g.,][]{Andre+2010}. In particular, results of the {\it Herschel} Gould Belt Survey (HGBS) in nearby molecular clouds ($d <$ 500\,pc) indicate that molecular filaments have a typical inner width of $\sim $0.1\,pc \citep{Arzoumanian+2011,Arzoumanian+2019}, and that $\sim $75\% of prestellar cores form within dense, thermally supercritical filaments, with masses per unit length (hereafter line masses for short) $M_{\rm line}$ larger than $\sim $16\,$M_\odot {\rm pc^{-1}}$, equivalent to gas surface densities above $\sim $160\,$M_\odot {\rm pc^{-2}}$ (or $A_{\rm V} \sim $8) \citep{Konyves+2015}. 

These findings support a filament paradigm for low-mass star formation in two main steps (e.g., \citealt{Andre+2014}): (1) large-scale compression of interstellar material in supersonic magneto-hydrodynamic (MHD) flows generates a cobweb of molecular filaments in the ISM; (2) the densest filaments fragment into prestellar cores by gravitational instability 
near or above the critical line mass of nearly isothermal, cylindrical filaments \citep[cf.][]{Inutsuka+1997}, $M_{\rm line, crit} = 2\, c_{\rm s}/G$ ($\sim $16\,$M_\odot {\rm pc^{-1}}$ for molecular gas at $\sim $10\,K). 
While the details of this filament scenario remain actively debated, there is little doubt that dense molecular filaments are representative of the initial conditions 
of at least low-mass star formation \citep{Hacar+2023,Pineda+2023}. 
Since real filaments are not straight, uniform cylinders but exhibit bends and density fluctuations along their lengths, it is important to stress that it is the local rather than 
the average value of the mass per unit length of a filament that determines its ability to fragment \citep[cf.][]{Inutsuka2001,Heigl+2018}.

The {\it Herschel} observations also confirm the existence of a close relationship between the prestellar core mass function (CMF) and the stellar initial mass function (IMF) in the regime of low to intermediate stellar masses ($\sim $0.1--5$\,M_\odot $ -- \citealt{Konyves+2015,Marsh+2016,Di+2020}), and suggest a connection between the CMF/IMF and the filament line mass function or FLMF \citep{Andre+2019}. Indeed, while the mass function of molecular clouds and clumps within clouds is significantly shallower than the Salpeter IMF ($\Delta N / \Delta \log M_{\rm cl} \propto M_{\rm cl} ^{-0.6\pm0.2}$ for clouds/clumps vs. $\Delta N / \Delta \log M_\star \propto M_\star^{-1.35}$ for stars above $1\,M_\odot $ -- e.g., \citealt{Salpeter1955,Solomon+1987,Kramer+1998}) the FLMF derived from  {\it Herschel} data for the sample of HGBS filaments analyzed by \citet{Arzoumanian+2019} is much more similar to the Salpeter IMF, with $\Delta N / \Delta \log M_{\rm line} \propto M_{\rm line}^{-1.6\pm0.1}$ above the thermally critical line mass of ${\sim 16}\,M_\odot {\rm pc^{-1}}$ \citep{Andre+2019}. This suggests that molecular filaments may represent the key evolutionary step during the non-linear growth of interstellar structures at which a Salpeter-like mass function is established.

The California giant molecular cloud (California GMC) belongs to the Taurus-Auriga-California-Perseus complex (TACP), as illustrated in Fig.~\ref{largemap}. The four giant molecular clouds on the right-hand side surround a region devoid of dust, CO, and HI \citep{Lim+2013}. That void is filled, however, with the $\rm H{\alpha}$ emission from hydrogen atoms ionized by massive stars \citep{Finkbeiner2003}. 
The \object{California GMC} is located at the front end of a HI supershell \citep{Shimajiri+2019a}, and its large-scale morphology, revealed by $^{13}$CO emission, is dominated by a continuous filamentary shape over $> 70$\,pc \citep{Guo+2021}.
The supernova explosions in the Per OB2 association may have created this HI supershell and pushed the gas forward \citep{Sancisi+1974,Hartmann+1997}. 
The \object{California GMC} was named ``Auriga'' in the GouldsBelt \emph{Spitzer} Legacy Survey \citep{Allen+2006}, and therefore also named ``Auriga-California'' by \cite{Harvey+2013} and \cite{BroekhovenFiene+2014}. 
However, \cite{Lada+2009} derived a distance of 450\,pc for this molecular cloud through a comparison 
of foreground star counts with Galactic models, which makes a very substantial difference with the distances of Taurus-Auriga (150\,pc) and Perseus (240\,pc). For this reason, this paper follows \citet{Lada+2009} 
in using the name ``California'' for this molecular cloud. \cite{Schlafly+2014} found a distance of $410\pm41$\,pc using optical photometry of stars observed by PanSTARRS-1 along the lines of sight toward this cloud. 
Considering that \emph{Gaia} measured distances with an accuracy never achieved before, we adopt a distance of $d = 470$\,pc in this study \citep{Zucker+2019}.

While the mass and plane-of-sky morphology, sizes, and kinematics of the \object{California GMC} are similar to those of the Orion~A GMC, the star formation rate observed in California is more than an order of magnitude lower than in Orion~A \citep{Lada+2009}. This difference in star formation activity has been interpreted as resulting from a difference in dense gas fraction by \citet{Lada+2009}. Alternatively, using $\it Gaia$ data, \citet{RezaeiKainulainen2022} recently pointed out that the 3D shape of California is very different from 
the large-scale 3D filamentary shape of Orion~A and corresponds to a sheet-like structure, extended by up to 120\,pc along the line of sight when viewed (almost edge-on) from the Sun, but with a depth of only $\la 25\,$pc when viewed face-on. They proposed that these differing 3D morphologies may partly account for the marked difference in star formation rate. 

Here, we use available {\it Herschel} data to present a detailed study of the global properties of the population of dense cores and filaments in the California GMC. 
The paper is organized as follows. Section~\ref{sec:observations} describes the {\it Herschel} observations and the derivation of column density images. 
In Sect.~\ref{sec:results}, we analyze the mass distribution within the cloud, extract dense cores and filamentary structures from the {\it Herschel} data, and 
derive the main associated statistical distributions, including the CMF and, for the first time the distribution of filament {\it local} line masses or local FLMF. 
Section~\ref{sec:discussion} discusses the results, emphasizing the likely physical connection between the local FLMF and the CMF in the California GMC, 
and revisiting an empirical toy model for the origin of the CMF/IMF in filaments. 
We summarize and conclude the paper in Sect.~\ref{sec:conclusions}. 


\begin{figure*}
   \centering
   \includegraphics[width=1.0 \hsize]{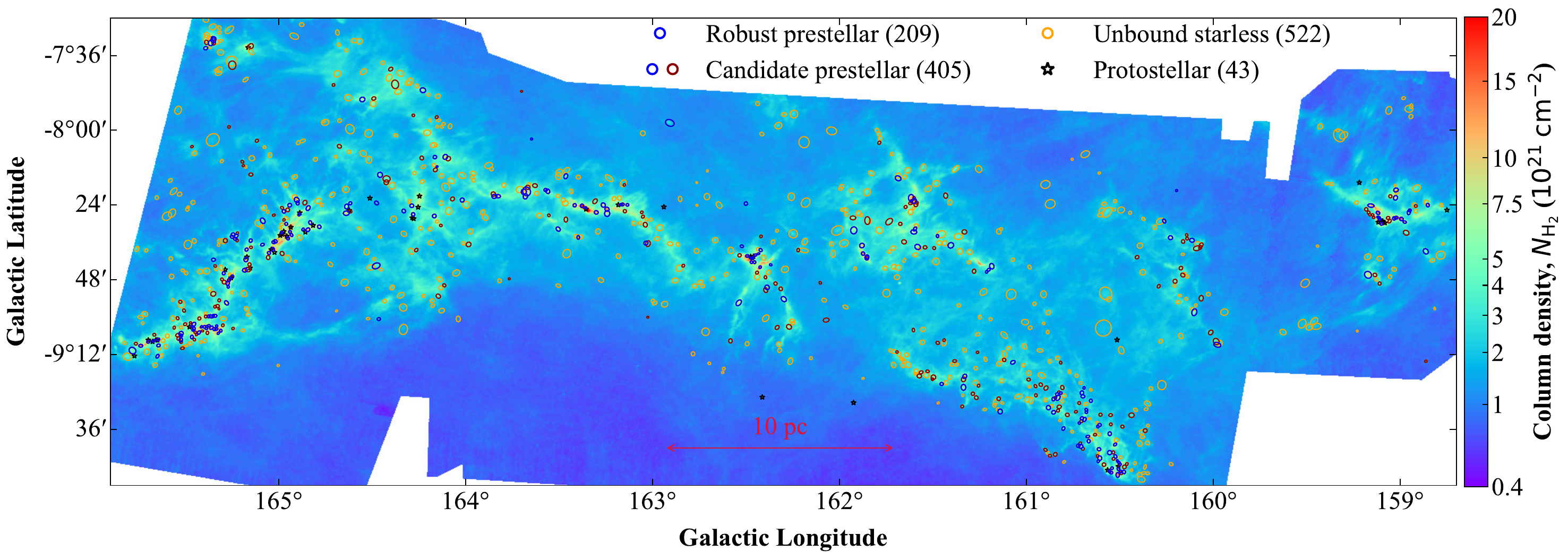}
   \caption{Column density map of the \object{California GMC} at an angular resolution of 18.2{\arcsec}, derived from {\it Herschel} data with \textsl{hires} (see Sect.~\ref{sec:observations}). 
The positions of the dense cores identified with \textsl{getsf} (see Sect.~\ref{selection_core}) are overlaid and shown as ellipses of various colors according to core type 
(blue for robust prestellar cores, blue and brown for candidate prestellar cores, orange for unbound starless cores). 
For easier visibility, the full axes of the ellipses have twice the actual Gaussian full width at half maximum (FWHM) sizes. 
Protostellar cores are marked with black pentagrams.}
\label{California18p2}
\end{figure*}

\section{{\it Herschel} data and derived column densities}\label{sec:observations}

The {\it Herschel} imaging observations of the \object{California GMC} \citep{Harvey+2013} were taken in parallel mode 
in the PACS 70 and 160~$\mu $m \citep{Poglitsch+2010} and SPIRE 250, 350, and 500~$\mu $m \citep{Griffin+2010} wavebands, with full width at half maximum (FWHM) 
beam sizes of 8.4, 13.5, 18.2, 24.9, and 36.3\arcsec, respectively, 
and a scanning speed of 60\arcsec s$^{-1}$. 
The corresponding {\it Herschel} multiwavelength data were downloaded from the NASA/IPAC Infrared Science 
Archive\footnote{\url{https://irsa.ipac.caltech.edu/data/Herschel/ACMC/}}.

To create a column density map at $18.2{\arcsec}$ resolution, we used the \textsl{hires} utility \citep{Menshchikov2021method}, an implementation within the \textsl{getsf} package of the multiresolution technique introduced by \citet[][see their Appendix A]{Palmeirim+2013}. 
The \textsl{hires} utility performs pixel-by-pixel fitting of the four {\it Herschel} images at 160, 250, 350, and 500~$\mu $m to derive dust temperature and column density images of the California GMC at various angular resolutions and uses the multi-scale differential approach of \citet{Palmeirim+2013} 
to improve the resolution of the resulting map to 18.2{\arcsec}.  

The derivation of reliable column densities and temperatures from {\it Herschel} data based on pixel-by-pixel fitting of the observed spectral energy distributions (SEDs)
requires that the input images have accurate low-level intensities on large scales (also known as ``zero levels'').
Due to their finite sizes, the original {\it Herschel} maps usually do not have satisfactory zero levels and need to be corrected by 
adding appropriate zero-level offsets. The latter can be estimated by comparing smoothed versions of the {\it Herschel} images 
with corresponding low-resolution (5\arcmin) data from the {\it Planck} and {\it IRAS} all-sky surveys (cf. \citealp{Bernard+2010, Bracco+2020}). The estimated 
zero-level offsets for the \object{California GMC} at 160, 250, 350, and 500~$\mu $m are 4.7, 35.9, 17.8, and 5.8\,MJy\,sr$^{-1}$, respectively.

The observed images were reprojected to the same grid, covering a total area of $7\degr \times 2.3\degr$ with a pixel size of $3${\arcsec}.
The \textsl{hires} utility generated images of the $\rm H_{2}$ column density $N_{\rm H_{2}}$ and dust temperature $T_{\rm d}$ at resolutions of 18.2, 24.9, and 36.3{\arcsec} resolutions. 
In this process, a dust opacity law $\kappa_{\lambda } = 0.1 \times (\lambda /300\,\mu\mathrm{m})^{-\beta}$ cm$^{2}$\,g$^{-1}$ with a fixed $\beta=2$ \citep{Roy+2014} was assumed. 
The column density image at 18.2{\arcsec} resolution ($\sim$\,0.04\,pc at $d=470$\,pc)  
is instrumental to accurately detect and measure sources and filaments and reveals more details about the molecular cloud than the lower-resolution maps. 
The 18.2{\arcsec} column density map (see Fig.~\ref{California18p2}) is publicly available in fits format from the CDS.

Having prepared the images, we applied the \textsl{getsf} source and filament extraction method \citep{Menshchikov2021method} to separate the structural components of sources, filaments, and background and extract the sources and filaments in their respective images. The \textsl{getsf} extraction method is an improved version of its predecessors \textsl{getsources}, \textsl{getfilaments}, and \textsl{getimages} \citep{Menshchikov+2012,Menshchikov2013,Menshchikov2017}. For a detailed benchmarking and comparisons of the new and old methods, the reader is referred to \cite{Menshchikov2021benchmark}. 

\section{Results and analysis}\label{sec:results}
\subsection{Mass distribution}
The California GMC exhibits a hierarchical pattern, comprising a broad smooth background, web-like filamentary structures, and compact sources such as dense cores. 
The \textsl{getsf} extraction method, as described by \citet*{Menshchikov2021method}, segregates the structural components into their separate images. These separated components can be used quantitatively as they are guaranteed to reconstruct the original image through the addition of the components. Their respective masses can be estimated by summing up all pixel values in their individual column density maps:
\begin{equation}
M_{\rm comp} = \mu_{\rm H_{2}} m_{\rm H} \Delta^{2} \sum_{i} N_{{\rm H_{2}}i}.
\label{masses}
\end{equation}
Here, $\mu_{\rm H_{2}}=2.8$ represents the mean molecular weight per $\rm H_{2}$ molecule, $m_{\rm H}$ denotes the mass of a hydrogen atom, 
$\Delta$ denotes the pixel size, and $N_{{\rm H_{2}}i}$ corresponds to the column density of the given structural component at pixel $i$. 
The total mass of the California GMC within the {\it Herschel}-observed region (Fig.~\ref{California18p2}) is $M_{\rm tot} = 3.08 \times 10^{4}\,M_\odot$. 
The resulting filamentary structures contain $6500\,M_\odot$ (21\% of $M_{\rm tot}$), and the sources contain 550\,$M_\odot$ (1.8\% of $M_{\rm tot}$). The visual extinction ($A_{\rm V}$) is determined from the $\rm H_{2}$ column density using the relationship $N_{\rm H_{2}} = 0.94 \times 10^{21}\ A_{\rm V}$ mag \citep{Bohlin+1978}. 

\begin{figure}
   \centering
   \includegraphics[width=1.0\hsize]{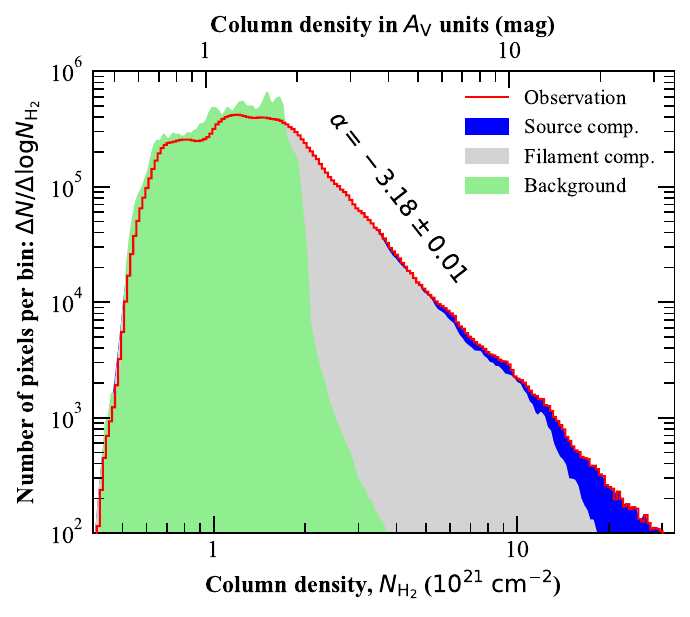}
\caption{Column density probability distribution function (PDF$_N$) obtained from the $18.2''$ resolution image shown in Fig.~\ref{California18p2},  along with its structural components. 
For $2 \leq A_{\rm V} \leq 5$, the PDF$_N$ follows a power law with an exponent of $-3.2$. It becomes slightly flatter for $5 \leq A_{\rm V} \leq 10$, but its power-law exponent returns to $-3.2$ 
for $A_{\rm V} > 10$.}
\label{PDF}
\end{figure}

The probability density function of column density PDF$_N$ is shown in Fig.~\ref{PDF}. The high-density tail of the PDF$_N$ exhibits a power-law shape suggestive of gravitational collapse \citep[e.g.,][]{FederrathKlessen2013}, with an exponent of $-3.2$, akin to the $-2.9$ value for Aquila  \citep{Konyves+2015} but steeper than the $-1.9$ value for Orion B \citep{Schneider+2013, Konyves+2020}. Notably, discernible source components emerge when $A_{\rm V} > 4$, becoming more pronounced at $A_{\rm V} > 6$. Defining dense gas as gas with $A_{V,\text{bg}} > $~4.5--7, the filament and source components correspond to 66--73\% and 10--14\% of the dense gas mass, respectively.

\subsection{Identification and analysis of dense cores}
\label{selection_core}

Following the standard approach introduced by \citet{Konyves+2015} for identifying dense cores in the {\it Herschel} images of nearby clouds taken as part of the HGBS, 
we performed two sets of \textsl{getsf} extractions -- tailored to starless cores (combining the 18.2{\arcsec} column densities with 250, 350, and 500~$\mu $m images for source detection) 
and to protostellar cores (detecting them only in the 70~$\mu $m image), respectively. Measurements of detected sources were done in all images.

From the \textsl{getsf} extraction catalogs, we selected only those sources that met the standard HGBS criteria for core identification, 
based on column density detection, 
global detection significance, goodness, monochromatic detection significance, goodness, and signal-to-noise ratio (see Sect.~4.5 of \citealt{Konyves+2015}).
Such a selection is crucial for eliminating potential spurious detections, ensuring that only reliable and well-measurable sources are included.
To further exclude extragalactic sources, we cross-matched the source positions with SIMBAD and the NASA/IPAC Extragalactic Database within a 6{\arcsec} matching radius, which left us with 970 acceptable candidate cores. 
Following the standard approach, sources that were detected at 70~$\mu $m and were, at most, marginally resolved ($\sqrt{AB} < 2 \times 8.4${\arcsec}), where $A$ represents the major axis size at half-maximum and $B$ represents the minor axis size at half-maximum, and were not too elongated ($A/B < 1.5$), were interpreted as candidate protostellar cores.
In the California GMC, this resulted in a sample of 
927 starless dense cores\footnote{The sensitivity of the {\it Herschel}/PACS observations of the California GMC was such that very-low-luminosity protostellar objects with 70~$\mu $m flux densities down to $\sim$100\,mJy 
\citep{Harvey+2013} and accretion luminosities down $< 0.1\, L_\odot $ \citep[cf.][]{Dunham+2008}
could be detected at the $\sim$\,$10\sigma$ level. Most (if not all) of the starless cores identified here, which lack significant emission in the PACS 70~$\mu $m data, are therefore truly starless.}
and 43 protostellar cores.

\begin{figure}
   \centering
   \includegraphics[width=1.0\hsize]{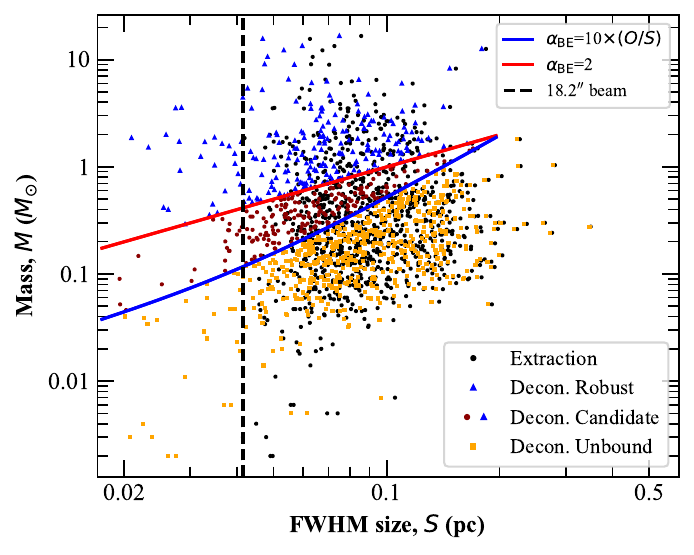}
\caption{Mass--size diagram for the cores identified in the California molecular cloud. "Extraction" represents the sizes measured directly on the 18.2{\arcsec} column density map. "Decon. Robust", "Decon. Candidate", and "Decon. Unbound" represent the sizes of robust prestellar, candidate prestellar, and unbound starless cores, respectively, after deconvolution. The red solid line represents half of the mass of a thermally critical Bonnor-Ebert sphere ($M_{\rm BE}$/2) of radius $R$ at a temperature of 10~K. The blue solid line marks the estimated threshold for selecting candidate prestellar cores.
The dashed vertical line marks the 18.2{\arcsec} beam size.}
\label{MassvsR}
\end{figure}

The FWHM size of a core was estimated as the geometric mean of its major and minor sizes, i.e.,  
$S = \sqrt{AB}$. We applied a simple Gaussian deconvolution method to remove the effect of the beam and recover a better estimate of the true source size from the observed $S$. 
Accordingly, the deconvolved size was estimated as $S_{\text{dec}} = \sqrt{S^2 - O^2}$, where $O$ corresponds to the 18.2{\arcsec} angular resolution of the column density map. 
The deconvolded sizes of sources with $S$ close to $O$ exhibit increased unreliability. 
To ensure reliable results and prevent unacceptably large deconvolution errors, we employed a resolvedness criterion and required $S/O > 1.1$ 
(see, e.g., \citealp{Menshchikov2023}). 
When $S/O < 1.1$, sources were considered partly resolved or unresolved, and assigned a deconvolved size equal to half of $S$. Figure~\ref{MassvsR} shows the mass versus size diagram for the starless cores identified in the California GMC. 
We use the thermal version of the critical Bonnor-Ebert mass, denoted as $M_{\text{BE}}^{\text{crit}}$, to identify candidate prestellar cores. 
Prestellar cores are those starless cores which are dense enough to be self-gravitating, 
making them strong candidates for future protostar formation. 
$M_{\text{BE}}^{\text{crit}}$ is an approximation of the virial mass for an unmagnetized, thermal core, and can be expressed as follows: 
$M_{\text{BE}}^{\text{crit}} \approx 2.4 R_{\text{BE}} c_{\text{s}}^2 / G$, 
where $R_{\text{BE}}$ is the Bonnor-Ebert radius,  $c_{\text{s}}$ is the isothermal sound speed $\sim $0.19\,km/s 
for an ambient cloud temperature of 10 K, and $G$ is the gravitational constant. 
We adopted $R_{\text{dec}} = S_{\text{dec}}\,d  $ as an estimate of $R_{\text{BE}}$.
Applying the criterion $\alpha_{\text{BE}} = M_{\text{BE}}^{\text{crit}}/M_{\text{core}} \leq 2$, we identified 209 starless sources as robust prestellar cores. 
Alternatively, applying the modified empirical formula  $\alpha _{\text{BE}} \le 10 \times (O/S$) based on our simulation tests  (see Appendix~\ref{corecomp}), 
we identified a larger sample of 405 candidate prestellar cores. 

\begin{figure}
   \centering
   \includegraphics[width=1.0\hsize]{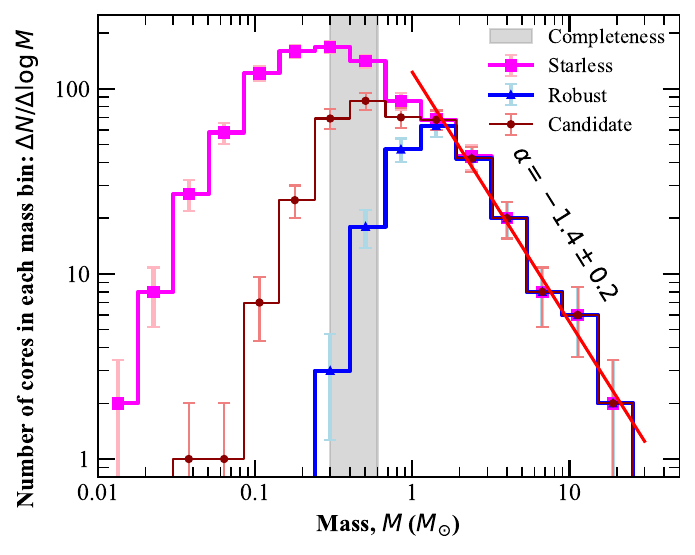}
\caption{Differential core mass function ($\Delta N/\Delta{\rm log}M$) for 927 starless cores (purple), 405 candidate prestellar cores and 209 robust prestellar cores. The vertical gray filled area at $0.45\pm0.15$\,$M_\odot$ denotes the estimated 90\% completeness level for prestellar cores (see Appendix~\ref{corecomp}). Error bars represent statistical uncertainties ($\sqrt{N}$). A power-law fit to the CMF at the high-mass end ($\gtrsim 1\,M_\odot$) yields a logarithmic slope of $-1.4\pm 0.2$ (see text).}
\label{CMF}
\end{figure}

The differential CMFs derived for the samples of starless cores, candidate prestellar cores, and robust prestellar cores are displayed in Fig.~\ref{CMF}. 
The three distributions merge at high masses (> 1\,$M_\odot$) and are consistent with a power-law mass function $\Delta N/\Delta {\rm log}M \propto M^{-\alpha}$ with $\alpha = 1.4 \pm 0.2$, which agrees very well with the Salpeter power-law 
IMF ($\alpha = 1.35$) \citep{Salpeter1955}.
A Kolmogorov-Smirnov (K-S) test on the cumulative core mass distribution $N(>M)$ confirms that it is statistically indistinguishable from a power-law distribution above 1\,$M_\odot$ at a \hbox{K-S} significance level  of $>$\,$40\%$ (equivalent to $0.9\sigma$ in Gaussian statistics). 
The error bar on the power-law index was estimated from the range of best-fit exponents for which the \hbox{K-S} significance level was larger than 5\% 
(equivalent to $2\sigma$ in Gaussian statistics) when comparing the observed CMF with a power law above a minimum core mass in the range 1.0--1.4\,$M_\odot $.

\begin{figure}
   \centering
   \includegraphics[width=1.0 \hsize]{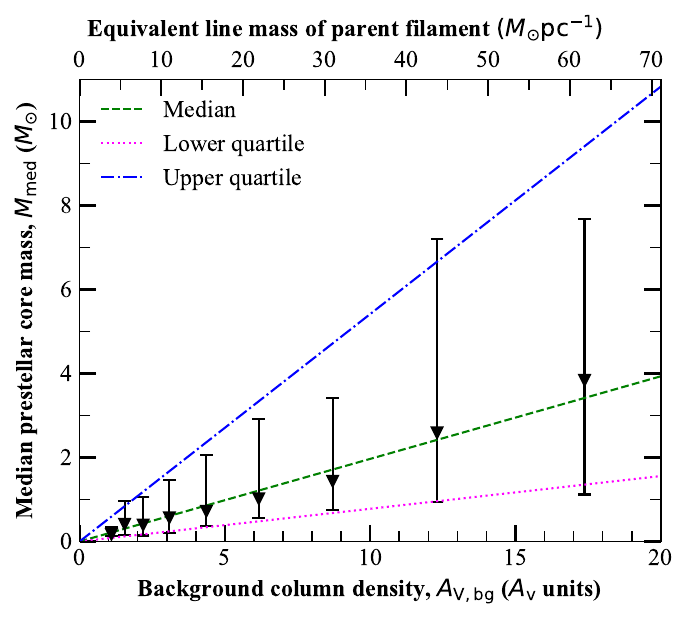}
\caption{Median candidate prestellar core mass plotted against background cloud column density ($A_{\rm V}$ units) is shown by the black triangles. The error bars represent the inter-quartile range of observed masses within each bin. The green dashed line depicts a linear fit to the median masses as a function of $A_{\rm V, bg}$, given by $M_{\rm med}\propto (0.2\pm0.02) \times A_{\rm V, bg}$. The blue dash-dotted line corresponds to a linear fit to the upper quartiles of median observed masses, while the magenta dotted line represents a linear fit to the lower quartiles. }
\label{massvsbg}
\end{figure}

Figure~\ref{massvsbg} shows the relationship between the median mass of prestellar cores and the local background level, which was estimated by subtracting the contribution of extracted sources from the original observed image while preserving the filament and large-scale background components. The two quantities exhibit a 
positive correlation, suggesting that more massive 
prestellar cores tend to form in denser parts of the ambient cloud 
(see \citealp{Konyves+2020,Shimajiri+2019b} for other examples of this trend). 

\subsection{Analysis of the filamentary network}
\label{sec:filaments}

Filaments within molecular clouds are elongated and relatively dense structures that are commonly found throughout these clouds (cf. \citealp{Menshchikov+2010,Andre2017,Arzoumanian+2019}). They exhibit an intricate network-like morphology, with interconnecting structures that form complex patterns. The physical parameters of the whole network of filamentary structures characterize the extent to which its parent molecular cloud collapses from a diffuse state to a dense structure (cf. \citealp{Ballesteros+2020}). The genesis of filaments within molecular clouds is believed to be a result of several physical processes, including large-scale turbulence, magnetic fields, 
and gravitational instability \citep[cf.][and references therein]{Andre+2014}. Filaments play a crucial role in facilitating the formation of dense cores by providing favorable conditions for the gravitational collapse of gas \citep[cf.][]{Li+2023}. Consequently, they serve as important nurseries for the birth of stars. 

\begin{figure*}
   \centering
   \includegraphics[width=1.0\hsize]{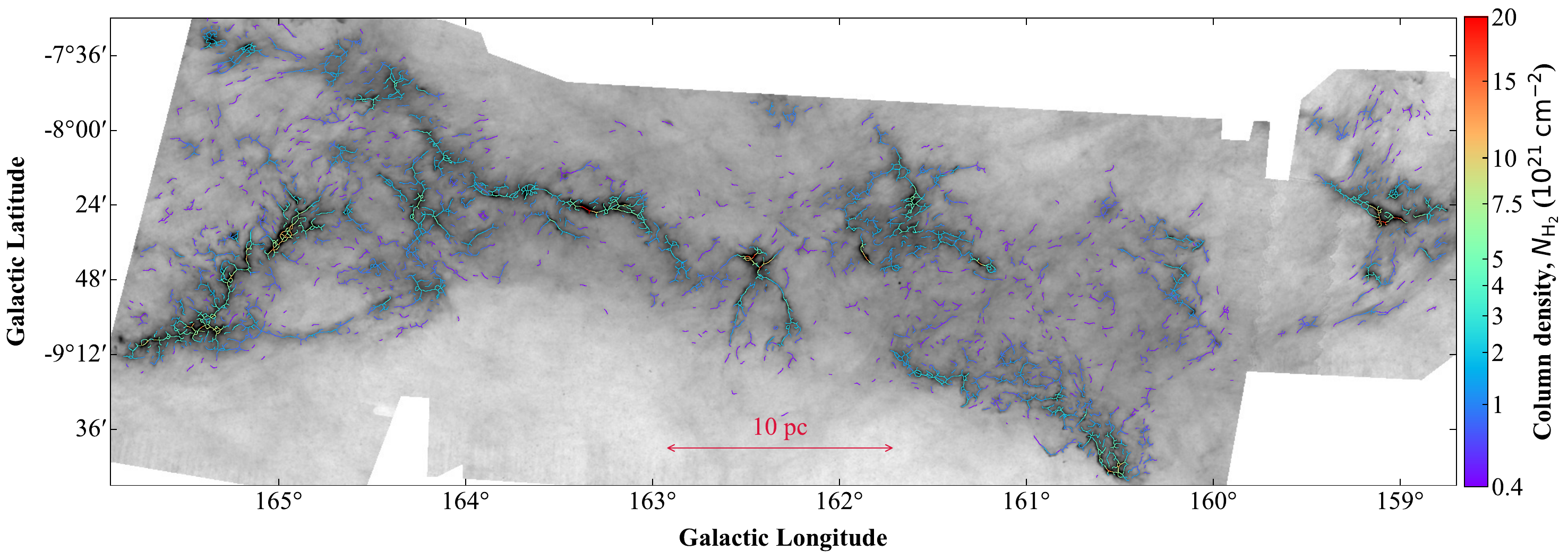}
\caption{Global filament network including all transverse angular scales up to $\pm$80\arcsec\ across the filament spines. 
A thick skeleton mask with a fixed line width of about 4 pixels (12\arcsec), color-coded as shown in the bar on the right, is overlaid on the original column density image shown as a grayscale background. Map values at the skeleton pixels correspond to background- and source-subtracted column densities on the filament crests.}
\label{network}
\end{figure*}

\begin{figure*}
\centering
    \includegraphics[width= 1.0\hsize]{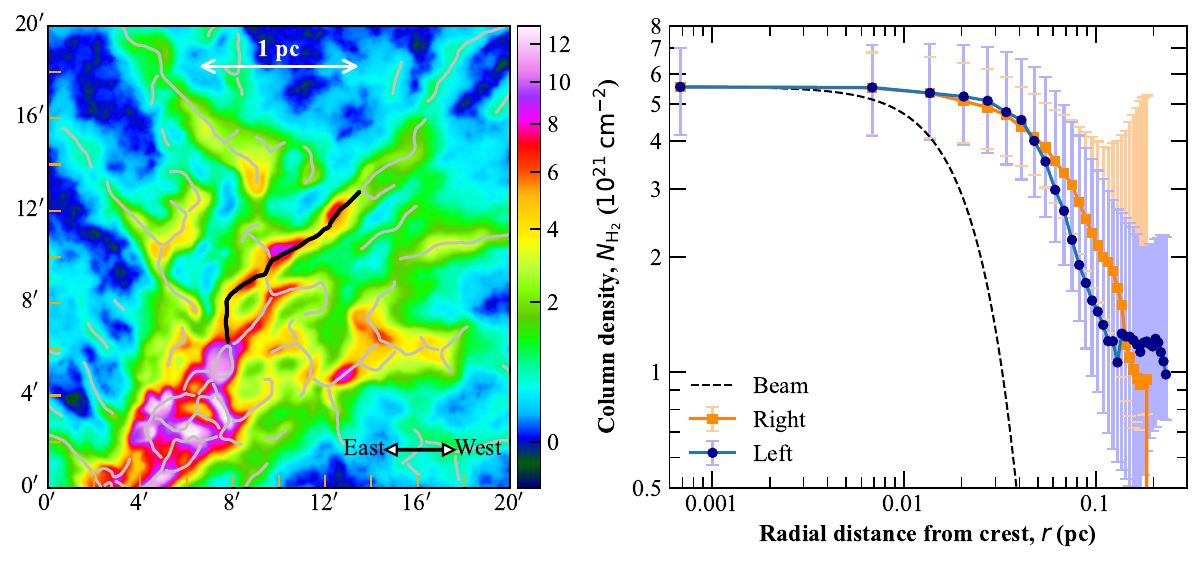}
\caption{Close-up view of the filament network (left) and example of a filament radial profile (right). 
{\bf Left:} Skeletons of several filaments depicted as grey solid curves and traced within a 20\arcmin\ $\times$ 20\arcmin\ subfield of the  column density image at 18.2{\arcsec} resolution. 
The black curve represents the skeleton of the filament whose radial profile is displayed in the right panel. 
{\bf Right:} Median one-sided column density profiles measured on both the left (purple solid line) and right (blue) sides of the filament, with the black dashed curve representing the beam profile. 
The median half maximum width measured on the left (north-east) side is 0.13\,pc, that on the right (south-west) side is 0.17\,pc, and the mean is 0.15\,pc.}
\label{filamentexample}
\end{figure*} 

In the present work, filaments were detected as part of the same \textsl{getsf} extraction process 
that we used to identify candidate dense cores in the {\it Herschel} data (see Sect.~\ref{selection_core}).
In our analysis, we chose a global skeleton, which traces filamentary structures on all transverse scales up to an angular diameter of 160\arcsec, i.e., approximately 0.4 pc, independently of the filament widths.
The corresponding filamentary network 
is illustrated in Fig.~\ref{network}, while a zoomed area of the detected filaments is presented in Fig.~\ref{filamentexample}. 
The filament skeleton map is publicly available in fits format from the CDS. 

While on large ($>10$\,pc) scales the California GMC appears to be a sheet-like 3D structure viewed almost edge on \citep{RezaeiKainulainen2022}, the vast majority 
of the smaller-scale ($\la1$\,pc) filaments and cores identified here are unlikely to be predominantly elongated along the line of sight as 1) this is statistically improbable for a large number of randomly-oriented substructures
embedded within a sheet-like cloud, and 2) several of our {\it Herschel} cores and filaments have been successfully observed and detected in dense gas tracers such as 
H$^{13}$CO$^+$(1--0) and N$_2$H$^+$(1--0) \citep{Zhang+2018,Chung+2019}, 
proving that they consist of dense $\ga 10^4\, {\rm cm}^{-3}$ gas as opposed to column density enhancements resulting from the accumulation of large amounts of low-density gas along the line 
of sight \citep[cf.][for the example of the Taurus main filament]{LiGoldsmith2012}.

\subsubsection{Measurement and statistics of extracted filaments} 
\label{filstat_sec}

\begin{figure*}
\centering
\includegraphics[width=1.0\hsize]{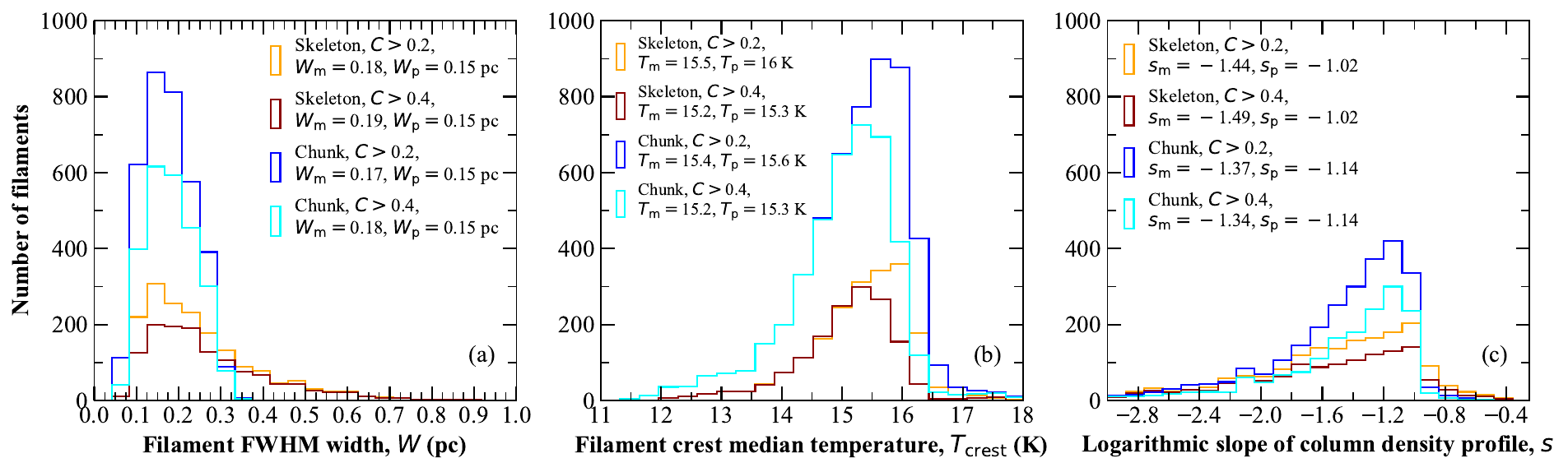}
\caption{
Distributions of widths, dust temperatures, and radial density gradients for filament skeletons and chunks (with $C > 0.2$ and $C > 0.4$). 
{\bf a)} Distribution of measured FWHM widths. 
{\bf b)} Distribution of median dust temperatures measured on filament crests. 
{\bf c)} Histogram of logarithmic slopes $s$ for the column density profiles of all measured filaments. 
(If both sides of a filament were effectively measured, the histogram count was incremented by 2;  if only one side was measured, the histogram count was incremented by 1; 
if neither side could be measured, the histogram count was not incremented.) 
The median and peak values of each parameter are provided in the legend of the corresponding panel.
}
\label{histogram}
\end{figure*}

Filamentary structures are more difficult to define and analyze than dense cores.  
In contrast to cores, observed filaments often have complex shapes, 
variable intensity along their crests \citep[cf.][]{Roy+2015}, 
and are frequently interconnected with each other 
\citep[e.g.,][]{Menshchikov2021method}. 
It is often unclear where a filament starts, where it ends, and which branch of the interconnected filamentary network describes a real physical filament. 
To enable an accurate analysis of the observations, it is necessary to decompose the complex filamentary pattern seen in the {\it Herschel} images
into simpler entities.
The approach employed by \textsl{getsf} is to break all branches of a filament network into simple non-branching structures, hereafter referred to as elementary skeletons, 
to aid in the selection of individual filamentary structures for which accurate measurements are possible \citep{Menshchikov2021method}. 

Properties of the filamentary network in the \object{California GMC} were measured in the column density map of the filamentary component at angular resolution $O = 18.2{\arcsec}$, free from the contributions of compact sources and large-scale backgrounds. The measurements were done with \textsl{fmeasure}, a utility of the \textsl{getsf} software. Both sides of the filaments were measured separately. For the purpose of estimating local filament properties, we also truncated each filament into segments of 0.1\,pc length, a size scale 
comparable to the typical half-power width of both HGBS and California filaments (see below and  \citealp{Arzoumanian+2019}). Such filament segments are referred to as ``chunks''  in the following (see also Sect.~\ref{FLMF_sec}).

Some filaments may be contaminated or blended with an adjacent filament; as a result, the width on one side or sometimes even both sides cannot be effectively measured. 
An effective measurement on one side is determined by two criteria: the profile falls below the half power maximum, and the width on this side is less than twice the width on the other side.
When estimating a filament's width, only the side(s) that could be effectively measured was considered. 
If both sides were effectively measured, the width $W$ was estimated as the mean of $W_{\rm A}$ and $W_{\rm B}$, $W=(W_{\rm A}+W_{\rm B})/2$, 
where $W_{\rm A}$ represents the filament's median half-maximum width measured on side $A$ and $W_{\rm B}$ is the measurement on side B. 
These widths are half-power diameters estimated by \textsl{getsf} directly from the profiles without any (Gaussian or Plummer) fitting, and 
are comparable to the ``half diameters'' $hd$ discussed by \citet{Arzoumanian+2019}. 
The distribution of observed widths for the filaments identified with \textsl{getsf} in California is shown in Fig.~\ref{histogram}\emph{a}. 

The column density contrast ($C$) of a filament may be defined as $C = N_{\rm H_{2},crest}/{N_{\rm H_{2},bg}}$, where $N_{\rm H_{2},crest}$ represents 
the median column density along the crest of the filament and $N_{\rm H_{2},bg}$ denotes the background column density of the filament \citep[cf.][]{Arzoumanian+2019}.
Higher-contrast filaments are easier to measure accurately due to a higher signal-to-noise ratio, 
since high contrast reduces noise impact and provides a clearer distinction from the background and other structures. 
The median value of $W_{\rm obs}$ for filament skeletons with $C > 0.4$ (0.19\,pc) deviates only slightly by 0.01\,pc from that for skeletons with $C > 0.2$ (0.18\,pc). Similarly, for filament ``chunks'', the median $W_{\rm obs}$ with $C > 0.4$ (0.18\,pc) also shows a 0.01\,pc deviation from that for chunks with $C > 0.2$ (0.17\,pc).
The agreement between the median filament widths obtained for $C > 0.4$ and $C > 0.2$
confirms the robustness of our measurements across different column density contrasts. 
The $W_{\rm obs}$ widths found here for the California GMC filaments  
are consistent within better than a factor of 2 with the typical 
inner width of $\sim $0.1\,pc 
measured for nearby ($d <0.5$\,kpc) molecular filaments 
with {\it Herschel} \citep{Arzoumanian+2011,Arzoumanian+2019}, especially if finite-resolution effects 
and the relatively large distance of the California GMC \citep[see][]{Andre+2022} are taken into account. 
Note that we do not try to deconvolve the measured filament widths from the telescope beam here, because the simple Gaussian deconvolution approach has been shown 
to be highly inaccurate, especially for filaments with power-law intensity profiles in the presence of significant background fluctuations 
\citep{Andre+2022,Menshchikov2023}.

The distribution of median temperatures along the filament crests is shown in Fig.~\ref{histogram}b. When $C > 0.4$, skeletons and chunks have median dust temperatures of 15.2\,K. The median dust temperatures for skeletons and chunks with $C > 0.2$ are only slightly warmer, at 15.5\,K and 15.4\,K, respectively. 

The distribution of measured logarithmic slopes $s$ for all filament column density profiles is presented in Fig.~\ref{histogram}\emph{c}, 
where the slope $s$ is defined as $s \equiv \mathrm{d} \ln N_{\mathrm{H}_2}/\mathrm{d} \ln r < 0$ (dimensionless) and measures how fast column density 
drops with radial distance from each filament's crest. 
Note that this drop is often quantified in terms of the logarithmic slope $-p = s-1$ of the underlying density profile, e.g., when Plummer-like model profiles are fitted to the data  
\citep[e.g.,][]{Arzoumanian+2011,Arzoumanian+2019}. 
The magnitude $|s|$ of the logarithmic slope directly controls the prominence of filament wings and the resemblance of the filament profile to a Gaussian shape. 
A higher $|s|$ (or $p$) value indicates less prominent filament wings and a filament profile that more closely resembles a Gaussian curve. 
The distribution of $s$ slopes has a peak at $-1$, and a median value at about $-1.45$ for filament skeletons and $-1.35$ for filament chunks. 
Accordingly, the $p$-index has a mode value of $+2$, and a median value of $+2.45$ for skeletons and $+2.35$ for chunks. 
Tests on high-contrast filaments (Contrast: $C>1$ and aspect ratio: $AR>4$) confirm that the logarithmic slopes derived by \textsl{getsf} (without Plummer fitting) are consistent with the $p$-index that would be derived by Plummer fitting.
Indeed, our results are in good agreement with the $p$-index slopes found by \citet{Arzoumanian+2019} for nearby HGBS filaments using Plummer model fits.

\begin{figure*}
\centering
\includegraphics[width=1.0 \hsize]{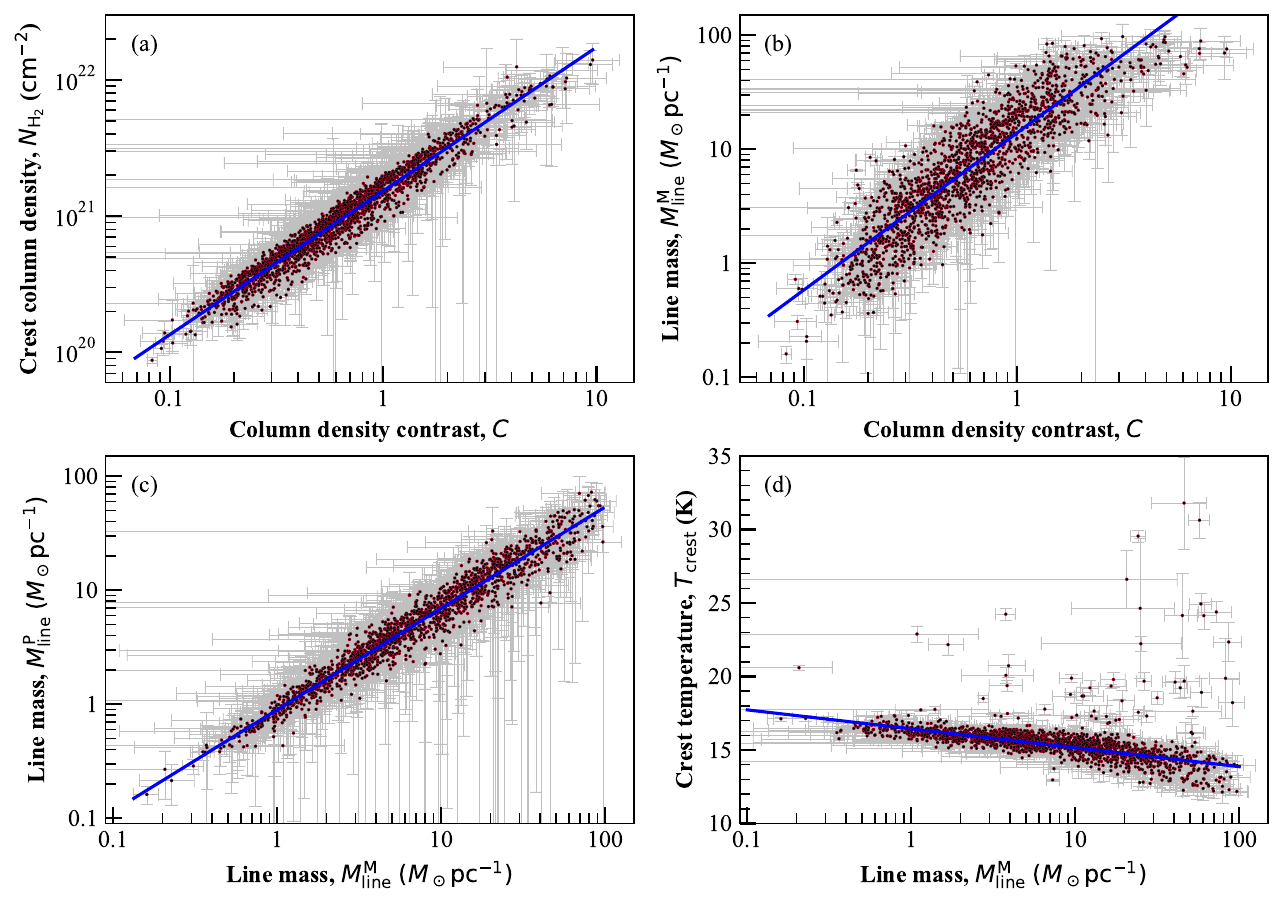}
\caption{Correlations between filament properties. {\bf (a)} Filament crest column density against column density contrast. The blue solid line is the fit: ${\rm log}_{10}(N_{\rm H_{2}})=1.06{\rm log}_{10}C+21.18$.  
{\bf (b)} Filament line mass ($M_{\rm line}^{\rm M} = M_{\rm fil} / L_{\rm fil}$)
versus column density contrast, well represented by the fit ${\rm log}_{10}M_{\rm line}^{\rm M}=1.38{\rm log}_{10}C+1.14$ (blue solid line). 
{\bf (c)} $M_{\rm line}^{\rm M}$ estimate against $M_{\rm line}^{\rm P}$ estimate of filament line mass (see text). The blue solid line is the fit: ${\rm log}_{10}M_{\rm line}^{\rm P}=0.89{\rm log}_{10}M_{\rm line}^{\rm M}-0.05$. 
(d) Filament crest temperature versus filament line mass $M_{\rm line}^{\rm M}$. The blue solid line represents the relation: $T=-1.29{\rm log}_{10}M_{\rm line}^{\rm M}+16.44$.}
\label{CNM}
\end{figure*}

Filament crest column densities and line masses are positively correlated with 
filament contrasts, as shown in Fig.~\ref{CNM}\emph{a,b}. This correlation indicates that higher-contrast filaments generally have higher column densities.
The average line mass of a filament, denoted as $M_{\rm line}^{\rm M}$, was estimated as $M_{\rm line}^{\rm M} = M_{\rm fil} / L_{\rm fil}$, where $M_{\rm fil}$ represents the filament mass, and $L_{\rm fil}$ corresponds to the filament length. The filament mass, $M_{\rm fil}$, was obtained by integrating column density over the entire footprint of the filament, while the length $L_{\rm fil}$ was estimated from the length of the \textsl{getsf} skeleton.
$M^{\rm P}_{\rm line}$ represents the line mass estimated from the integration of the median column density profile. $M^{\rm P}_{\rm line}$ is positively correlated with $M^{\rm M}_{\rm line}$, as shown in Fig.~\ref{CNM}\emph{c}. However, $M^{\rm M}_{\rm line}$ tends to be slightly higher than $M^{\rm P}_{\rm line}$. 
For high-contrast filaments ($C > 0.4$), $M_{\rm line}^{\rm M}$ appears to be a slightly better estimator of filament line mass than $M_{\rm line}^{\rm P}$ (cf. Appendix~B). 

Higher line-mass filaments tend to have lower dust temperatures
as shown in Fig.~\ref{CNM}\emph{d}. As the density of a filament increases, resulting in a higher line mass, the filament becomes more effectively shielded against ambient radiation, leading to a decrease in filament dust temperature.

\subsubsection{Relation between filaments and cores}

\begin{figure}
   \centering
   \includegraphics[width=1.0\hsize]{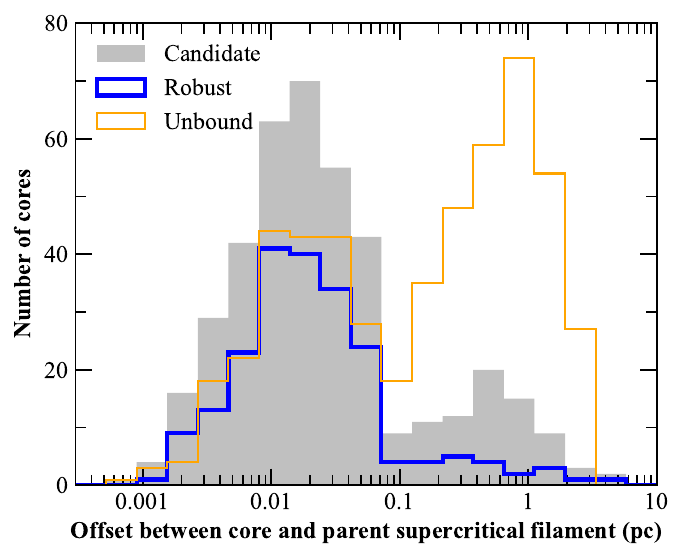}
\caption{Spatial relationship between extracted cores and supercritical filamentary structures, displayed as an histogram of separations between 
core centers and the nearest supercritical filamentary structures.} 
\label{offset}
\end{figure}

Filament line masses are known to vary along the filament crests \citep[cf.][]{Roy+2015}. 
Here, the line masses were estimated at each sampling point along the crests. 
For each core, the shortest separation between core center and the local network of supercritical filamentary structures was calculated. As shown in Fig.~\ref{offset}, our observations reveal that 82\% of candidate prestellar cores and 90\% of robust prestellar cores are positioned within 0.1\,pc of the nearest supercritical filament. In contrast, only approximately 41\% of unbound starless cores are situated within 0.1\,pc of the nearest supercritical filament. These results are consistent with the findings of \citet{Konyves+2020} and  \citet{Ladjelate+2020}, and support the hypothesis that filamentary structures act as pathways for the accumulation of gas and dust, leading to supercritical fragmentation and the subsequent formation of gravitationally bound cores. A majority ($\sim$60\%) of unbound starless cores and a small ($\sim$10--15\%) fraction of prestellar cores are located far ($\sim$\,0.1\,pc) from any supercritical filament, 
making up the secondary peak in the core-to-filament separation plot of Fig.~\ref{offset}. The nature of the cores in this second peak is not completely clear. 
Most of the unbound cores may never get massive enough to collapse and form stars before dispersing and may thus be ``failed cores'' \citep[cf.][]{Vazquez2005}. 
On the other hand, the few candidate prestellar cores populating the second peak in Fig.~\ref{offset} may be the progenitors of a small fraction of protostars that do not form in filaments.

\subsubsection{Line mass function of California filaments}
\label{FLMF_sec}

Background and noise fluctuations in observed images can easily break a continuous filament into multiple shorter filaments. As a result, when estimating integral quantities, such as the filament mass and length, the presence of unrelated fluctuations and the relative brightness (density) of the filament can introduce inaccuracies. Consequently, the physical conditions required for the formation of prestellar cores, particularly the volume densities, can vary along the filaments. This variability in physical conditions poses challenges when attempting to utilize the integral properties of filaments in studies related to star formation. However, it is not crucial to accurately determine the true integral properties of filaments. Instead, what matters more for star formation studies are the local properties of filaments in the immediate vicinity of the locations where appropriate physical conditions exist for the growth of prestellar cores.

Long filaments often exhibit significant variations in their local column densities and widths along their crests, which correspond to variations in their line masses denoted as ${M}_{\rm line}(l)$ \citep{Roy+2015}, where $l$ represents the position along the skeleton of a filament. 
We employed the \textsl{fmeasure} routine from the \textsl{getsf} software  
to trace each filament skeleton and determine the coordinates of all its pixels. 
To enhance the smoothness of the skeleton representation, \textsl{getsf} averages the coordinates of the pixels of a given skeleton over a seven-pixel window. This averaging process yields high-resolution coordinates for each point along the skeleton. 
Next, utilizing these high-resolution coordinates, each filament was truncated into pieces of 0.1\,pc length, which is comparable to the typical half-maximum width of both HGBS and California filaments (see \citealp{Arzoumanian+2019} and Sect.~\ref{filstat_sec}). This truncation results in the creation of equally-spaced filament segments referred to as ``chunks'' here. Consequently, a filament network composed of individual 0.1\,pc chunk units is established. 

\begin{figure}
\centering
\includegraphics[width=1.0 \hsize]{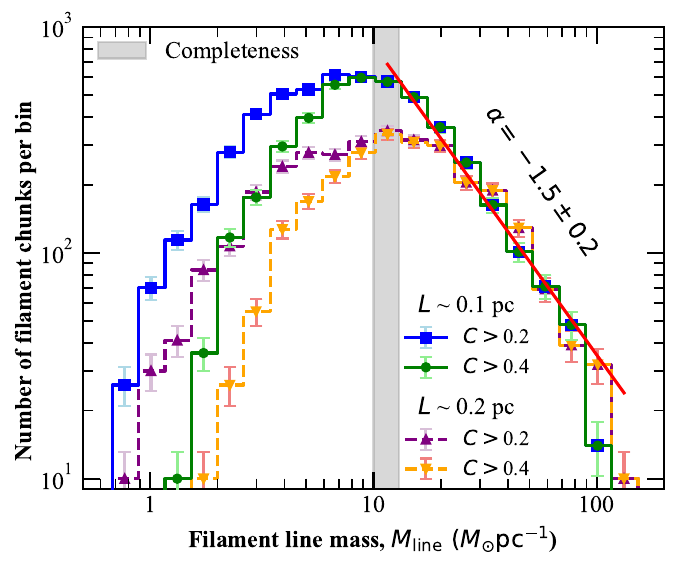}
\caption{
Filament line mass function (FLMF, $\Delta N/\Delta{\rm log}M_{\rm line}$), derived from the sample of filament chunks in the California GMC. 
The error bars represent statistical uncertainties ($\sqrt{\Delta N}$) in the numbers of filament chunks in each bin. The FLMF based on filament chunks with length $L \sim 0.1$\,pc and contrasts $C>0.2$ are shown by the blue histogram; the FLMF of chunks with $L \sim 0.1$\,pc and $C>0.4$  by the green histogram.
These two FLMFs coincide for ${M}_{\rm line} > 8\ M_\odot{\rm pc^{-1}}$. 
Above an estimated completeness limit of ${M}_{\rm line} \approx 10\ M_\odot{\rm pc^{-1}}$
(gray filled area), the filament chunk sample is considered 80\% complete (see Appendix~B). 
The FLMF can be fitted by a power law above $10\ M_\odot{\rm pc^{-1}}$ with an index of $-1.4\pm0.1$ (red solid line).
For comparison, the FLMF of filament chunks with $L \sim 0.2$\,pc and $C > 0.2$ is shown as a purple dashed histogram; the FLMF of chunks with $L \sim 0.2$\,pc and $C > 0.4$ 
as an orange dashed histogram.
} 
\label{FLMF}
\end{figure}

The ${M}_{\rm line}(s)$ values measured for each of the 0.1\,pc chunks were used to construct the filament line mass function (FLMF) shown in
Fig.~\ref{FLMF} as blue and green histograms.
We stress that since the 0.1\,pc length of each filament chunk significantly exceeds the linear resolution of our column density map (18.2{\arcsec} or $\sim $0.04\,pc at $d = 470$\,pc), the measured ${M}_{\rm line}(s)$ values are 
mutually independent. 
As described in Appendix~\ref{filamentcompleteness}, the derived sample of filaments is deemed to be 80\% complete for ${M}_{\rm line} >10\,M_\odot\,{\rm pc^{-1}}$. 
Figure~\ref{FLMF} shows that, in this regime of high line masses,
${M}_{\rm line} > 10\,M_\odot{\rm pc^{-1}}$, 
the FLMF is consistent with a power law: ${\Delta}N/{\Delta}{\rm log} {M}_{\rm line} \propto {M}_{\rm line}^{-1.4\pm0.1}$.

A Kolmogorov-Smirnov (K-S) test on the cumulative line mass distribution $N(>M_{\rm line})$ confirms that it is statistically indistinguishable from a power-law distribution ${\Delta}N/{\Delta}{\rm log} {M}_{\rm line} \propto {M}_{\rm line}^{-1.7\pm0.1}$ at the K-S significance level  of 10\% (equivalent to $1.6\sigma$ in Gaussian statistics) in the regime of thermally supercritical filaments, i.e., for ${M}_{\rm line} > 16\,M_\odot {\rm pc^{-1}}$. 
The error bar on the power-law index was estimated from the range of best-fit exponents for which the \hbox{K-S} significance level was larger than 5\% 
(equivalent to $2\sigma$ in Gaussian statistics). 
Comparing the best-fit power laws to the FLMF observed above a minimum ${M}_{\rm line}$ value in the range from 10 to 25~$M_\odot {\rm pc^{-1}}$, 
we conclude that the power-law index  of the FLMF is $1.5 \pm 0.2$ in the regime of high line masses. 
For comparison, each filament was also truncated into pieces of 0.2\,pc in length, generating a filament network consisting of 0.2\,pc chunks. 
The corresponding FLMF is also displayed in Fig.~\ref{FLMF} as purple and orange dashed histograms.
The latter is essentially identical to the 0.1\,pc chunk FLMF for line masses $> 20\,M_\odot{\rm pc^{-1}}$.

\section{Discussion: Connecting the FLMF and the CMF}\label{sec:discussion}

The shape of the California FLMF found in Sect~\ref{FLMF_sec} and shown in Fig.~\ref{FLMF} is consistent with  
the \citet{Salpeter1955} power-law IMF at the high-mass end. This contrasts with the mass function of molecular clouds and massive clumps within clouds, which is known to be significantly shallower than the Salpeter IMF, scaling as $\Delta N/\Delta {\rm log}{M}_{\rm cl }\propto M_{\rm cl}^{-0.6\pm0.2}$ \citep[e.g.,][]{Solomon+1987,ELada+1991,Blitz+1993,Ellsworth+2015,Rice+2016}. The finding of Sect~\ref{FLMF_sec} 
strongly supports  the existence of a connection between the FLMF and the CMF/IMF, as proposed by \citet{Andre+2019} from an analysis of about 600 filaments in 8 nearby clouds covered by the HGBS \citep[cf.][]{Arzoumanian+2019}. Our present result shows that this connection holds at the level of an individual cloud, namely California. Moreover, the FLMF derived by \citet{Andre+2019} was a distribution of mean line masses averaged along the length of each sampled filament, while the FLMF shown here in Fig.~\ref{FLMF} is a distribution of local line masses for a complete, homogeneous sample of filaments in the same cloud. Since it is the local line mass of a filament which defines its ability to fragment at a particular location along its spine and not the average line mass of the filament \citep[cf.][and references therein]{Pineda+2023}, the connection found here is more direct and provides tighter constraints on the origin of the CMF/IMF. 
Combining their result on the FLMF of HGBS filaments with the trend that higher-line-mass filaments tend to host more massive cores \citep[see][and Fig.~\ref{massvsbg} here]{Shimajiri+2019b,Konyves+2020}, \citet{Andre+2019} proposed an empirical model for the origin of the CMF/IMF in filaments, whereby the global prestellar CMF -- and by extension the global stellar IMF -- arises from the superposition of the CMFs generated by individual filaments \citep[see also][]{Lee+2017}. Based on average line masses, the model had to assume that HGBS filaments have fixed, typical lengths of $\sim $0.55\,pc and uniform line masses along their lengths, which is obviously only a first order approximation. 
The result reported here allows us to simplify and improve this model by dropping the artificial assumptions of constant filament length and uniform line mass along each spine.

\begin{figure}
   \centering
\includegraphics[width=1.0\hsize]{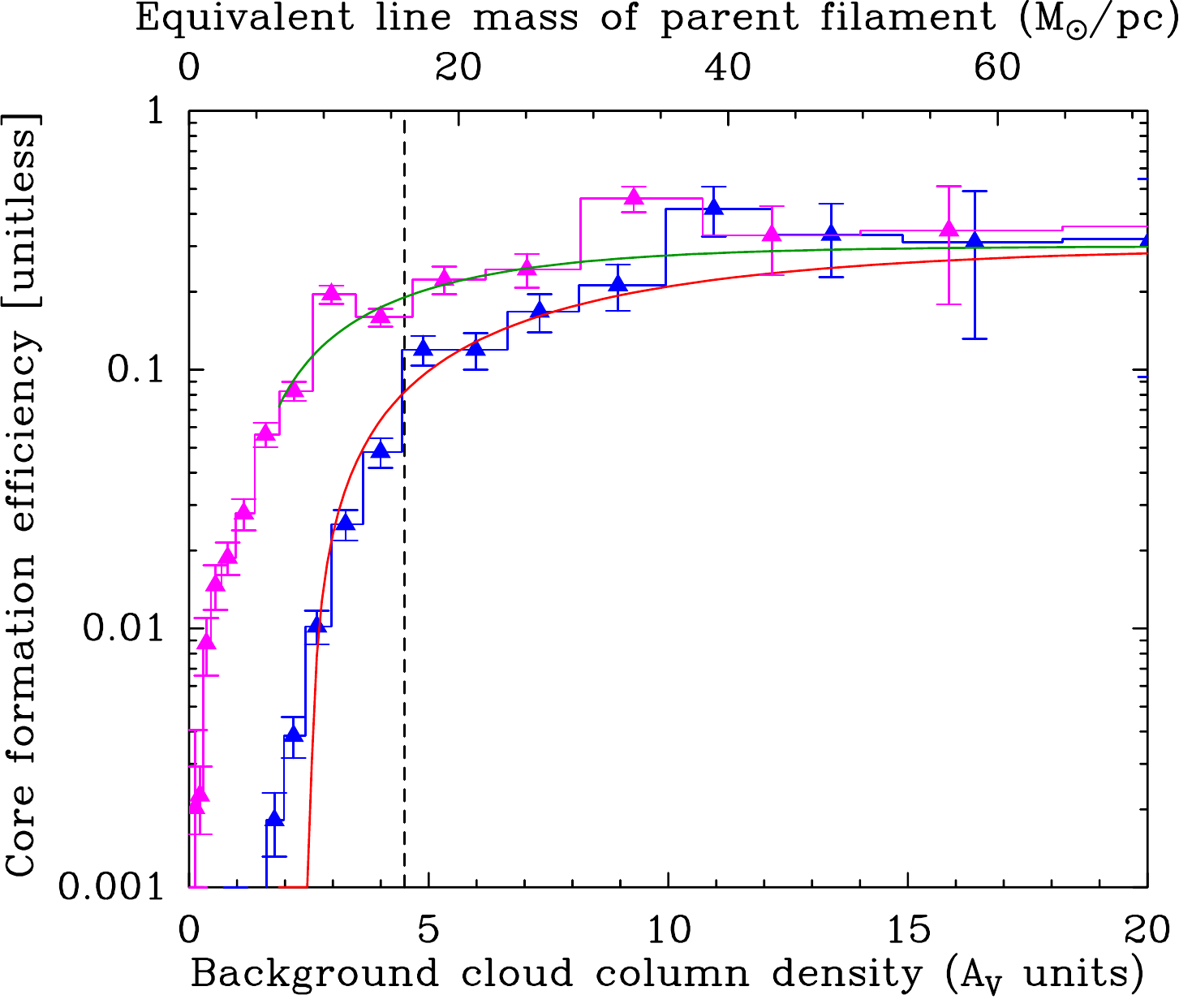}
\caption{Differential core formation efficiency (CFE) in the California GMC as a function of background cloud column density (blue histogram with error bars) 
and parent filament column density (magenta histogram),  both expressed in $A_{\rm V}$ units. 
The CFE values were obtained by dividing the mass in the form of candidate prestellar cores in a given column density bin by the cloud mass (blue histogram)
or estimated filament mass (magenta histogram) 
in the same column density bin. 
The upper x-axis provides a rough estimate of the equivalent filament line mass assuming 
the parent filament of each core has an effective width 0.18\,pc (corresponding to the median observed filament width -- see Sect.~\ref{filstat_sec}). The vertical dashed line marks the position of the thermally critical line mass $M_{\rm line, crit} \sim 16\,M_\odot\,{\rm pc^{-1}}$, corresponding to $A_{\rm V,\, bg} \sim 4.5$. 
The red and green curves show simple fits to the blue and magenta histograms, respectively, with a smooth step function of the form 
${\rm CFE}(A_{\rm V})  = {\rm CFE_{max}} [1- {\rm exp}(a\times A_{\rm V} + b)] $, with ${\rm CFE_{max}} \approx 30\% $. 
For the blue distribution, $a= -0.16$ and $b= 0.4$;  for the magenta distribution, $a= -0.28$ and $b= 0.25$. 
}
\label{cef}
\end{figure}

\begin{figure}
\includegraphics[width=1.0\hsize]{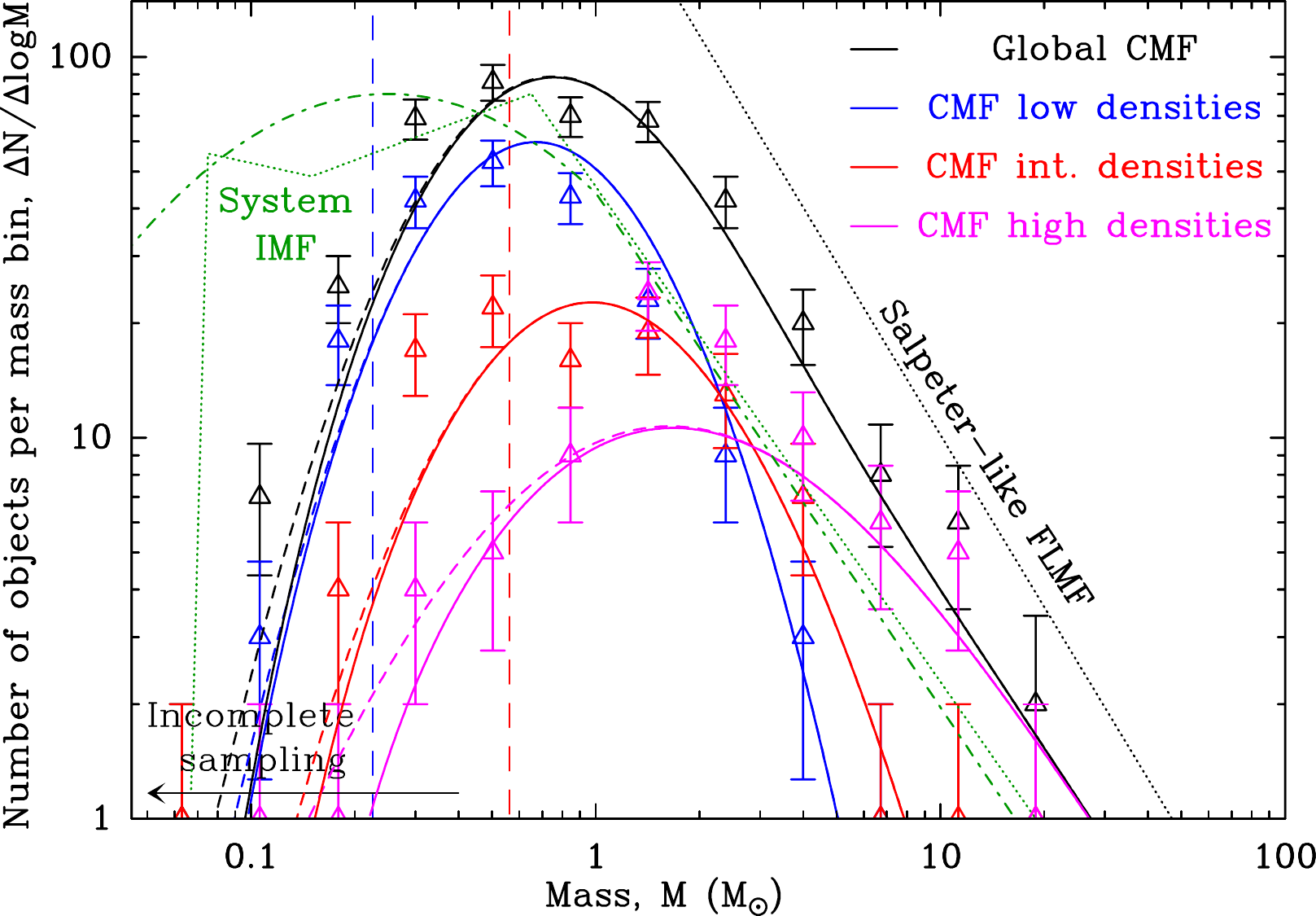}
\centering
\caption{
Comparison of the prestellar CMFs (in $\Delta N/\Delta$log$M$ format) expected in the toy model described in \citet{Andre+2019} and in 
the text (solid and dashed curves) with the observed prestellar CMFs observed in the \object{California GMC} (data points and histograms) 
at low column densities  ($2 < A_V^{\rm back} < 4.5$, in blue), 
intermediate column densities  ($4.5 < A_V^{\rm back} < 7$, in red), 
high column densities ($7 < A_V^{\rm back}$, in pink), 
and overall (all $A_V^{\rm back}$, in black).
(The solid curves represent the toy model including the estimated effect 
of incompleteness, while the dashed curves show the model ignoring the effect of incompleteness.)
The black dotted line displays the Salpeter-like power law FLMF, d$N$/dlog$M_{\rm line} \propto M_{\rm line}^{-1.5}$, 
assumed in the toy model and consistent with the observed FLMF in the supercritical regime (see Fig.~\ref{FLMF}).  
The green dash-dotted and dotted curves show the system IMFs from \citet{Chabrier2005} and  \citet{Kroupa+2013}, respectively.
The two vertical dashed lines mark the estimated $80\% $ completeness limits of the {\it Herschel} census 
of prestellar cores in California at low and high background column densities, respectively. 
Note how the CMF extends to higher masses at higher column densities, i.e., higher $M_{\rm line}$ filaments 
in both the toy model and the observations. }
\label{cmf_fil}
\end{figure}

Briefly, our revised model of the CMF may be summarized as follows (see \citealp{Andre+2019} for further details). 
Denoting by $ f_{M_{\rm line}}(m) \equiv {\rm d}N_{M_{\rm line}}/{\rm d\, log}\, m $ the differential CMF per unit log mass in a filament of line mass $M_{\rm line}$ 
(where $m$ represents core mass),  
and by $ g(M_{\rm line}) \equiv {\rm d}N/{\rm d\, log} \, M_{\rm line} $ the differential FLMF per unit log line mass, the model posits that, at least in a statistical sense, the global prestellar CMF per unit log mass  $\xi (m) \equiv {\rm d}N_{\rm tot}/{\rm d\, log}\, m $ in a molecular cloud corresponds to a weighted integration over line mass of the CMFs generated by the individual filaments present 
in the cloud\footnote{This is analogous to constructing the integrated galaxy-wide IMF (IGIMF) by 
summing the IMFs of all star formation events in the whole galaxy at a given time \citep[e.g.,][and references therein]{Jerabkova+2018}.}:
\begin{equation}
 \xi (m)  =   \int  f_{M_{\rm line}}(m)\, w(M_{\rm line})\, g(M_{\rm line})\, d {\rm log} M_{\rm line}, 
\end{equation} 

\noindent
where $w(M_{\rm line}) \propto {\rm CFE}(M_{\rm line}) \times M_{\rm line}  \times L$ represents 
the relative weight\footnote{Each filament segment is thus given a weight proportional to the total mass it harbors in the form of cores. 
The actual weights $w(M_{\rm line})$ applied in Eq.~(2) are normalized such as $\int  w(M_{\rm line})\,  d {\rm log} M_{\rm line} = 1$ and are therefore dimensionless.} 
of a filament segment of length $L$ (here $L=0.1$\,pc or $L=0.2$\,pc)
as a function of $M_{\rm line}$, and ${\rm CFE}(M_{\rm line}) \equiv M_{\rm cores}/(M_{\rm line}\, L)$ is the average prestellar core formation efficiency in a filament of line mass $M_{\rm line}$. 
Based on Fig.~\ref{cef} and the observational findings of \citet{Konyves+2015}, we adopt a dependence of 
CFE on $M_{\rm line}$ similar to \citet{Andre+2019}:
\begin{equation} 
{\rm CFE}(M_{\rm line})  = {\rm CFE}_{\rm max} \left[1\, -\, {\rm exp}\, \left(\frac{1}{4} - \frac{4\,M_{\rm line}}{3\,M_{\rm line, crit}}\right) \right],  
\end{equation} 

\noindent
with ${\rm CFE}_{\rm max} = 30\% $. 
This ``smooth step'' functional form for ${\rm CFE}(M_{\rm line})$ is such that the prestellar core formation efficiency in each filament asymptotically approaches 
${\rm CFE}_{\rm max} = 30\% $ for $M_{\rm line} > M_{\rm line, crit}/2$ and quickly drops to ${\rm CFE}(M_{\rm line})  = 0$ for $M_{\rm line} < M_{\rm line, crit}/2$ 
(see green curve and magenta histogram in Fig.~\ref{cef}). 
It provides a good fit to the core formation efficiency measurements in California (see Fig.~\ref{cef}), separating a regime with no core/star formation in thermally subcritical filaments from a regime with an asymptotic core formation efficiency of $\sim $30\% in thermally supercritical filaments. 
Likewise, based on the findings of \citet{Shimajiri+2019b,Shimajiri+2023} and \citet{Konyves+2020} and following \citet{Andre+2019}, we assume that the CMF produced by an individual filament, $f_{M_{\rm line}}(m)$, follows a lognormal distribution of variance $ \sigma_{M_{\rm line}}^2 $
centered on the effective Bonnor-Ebert (or Jeans) mass $ M_{\rm BE, eff}  \sim 1.3\, c_{\rm s, eff}^4 /(G^2 \Sigma_{\rm fil})$ in that filament, where $c_{\rm s, eff}$ is the one-dimensional gas velocity dispersion or effective sound speed and $ \Sigma_{\rm fil} \propto M_{\rm line}$ is the local gas surface density in the filament. Recalling that thermally supercritical filaments are 
observed to be virialized with $c_{\rm s, eff} \propto \Sigma_{\rm fil}^{0.5} \propto M_{\rm line}^{0.5}$ \citep[][see also \citealt{FiegePudritz2000}]{Arzoumanian+2013}, 
we see that $ M_{\rm BE, eff} \propto M_{\rm line}$. Thus, the model assumes that the location of the CMF peak in each filament scales with $M_{\rm line}$. 
In other words, higher-mass cores form in more massive filaments, a trend suggested by observations \citep[][and Fig.~\ref{massvsbg} here]{Shimajiri+2019b,Shimajiri+2023}. 
To reproduce the increasing dispersion of core masses with background column density (Fig.~\ref{massvsbg}), 
we adjusted the variances of the lognormal distributions $ f_{M_{\rm line}}(m)$ in the model as follows: 
$ \sigma_{M_{\rm line}}^2 = 0.3^2 +  0.2\, \left[{\rm log}\,(M_{\rm BE, eff}/M_{\rm BE, th})\right]^2 $.
The prestellar CMFs predicted by this empirical model in various column density regimes 
are shown in Fig.~\ref{cmf_fil}. 
It can be seen that the model reproduces both the global CMF and the variations of the CMF with background column density reasonably well. 

Our measurements of the FLMF and prestellar CMF in the California GMC (Sect.~\ref{sec:results}), 
coupled with the analysis and toy model presented in this section, therefore bolster the notion that the prestellar CMF -- and by extension the stellar IMF itself --
may be at least partly inherited from the FLMF. 
Interestingly, the numerical study by \citet{Abe+2021} suggests that only specific theoretical scenarios for filament formation and evolution, 
such as the oblique MHD shock compression mechanism \citep{Inoue+2013}, 
can account for a tight connection between the FLMF and the CMF/IMF, 
while other mechanisms, such as turbulence-induced shear or collision of turbulent shock waves, 
either do not produce massive enough filaments or 
do not lead to line mass functions resembling the Salpeter IMF at the high-mass end. 
The global hierarchical collapse (GHC) model advocated by \citet{Vazquez+2019} may potentially explain the link between 
the slope of the FLMF and the slope of the CMF/IMF since cores continuously accrete from their parent filamentary structures
and protostars similarly accrete from their parent cores in this picture. 
However, the difference in slope between the cloud/clump mass function and the slope of the FLMF is 
difficult to understand in the GHC framework since dense filaments are also observed to accrete 
from their parent clouds/clumps \citep[e.g.,][]{Palmeirim+2013, HKirk+2013, Shimajiri+2019a}. 
In any event, the FLMF appears to be a key feature of the population of dense molecular filaments 
that closely relates to the origin of stellar masses and deserves further attention from theorists of the cold ISM.

\section{Conclusions}\label{sec:conclusions}
We used {\it Herschel} SPIRE/PACS parallel-mode maps in five photometric bands between 70 and 500~$\mu $m to study 
dense cores and filamentary structures in the California molecular cloud, employing the \textsl{getsf} algorithm for source and structure identification. 
The physical properties of these cores and filaments were analyzed, and the relationship between the FLMF and 
the CMF discussed. 
Our main findings and conclusions can be summarized as follows:

   \begin{enumerate}
   \item The structure of the California GMC is hierarchical in nature, whereby the majority ($\sim $80\%) of the cloud's overall mass is distributed in 
   a large-scale, low-density background, $\sim $20\% of the mass is concentrated in the form of filamentary structures, and only $\sim $2\% of the mass 
   is in the form of compact sources such as dense cores.   
      
   \item We identified 43 protostellar cores and 927 starless dense cores, 405 of which were classified as candidate prestellar cores, including 209 robust prestellar cores. 
   For core masses $M > 1\,M_\odot$, the prestellar CMF is well fit by a power law $\Delta N/\Delta {\rm log}M\propto M^{-1.4 \pm 0.2}$,
   in excellent  agreement with the Salpeter power-law IMF. 
   
    \item A vast majority of robust (90\%) and candidate (82\%) prestellar cores are located within 0.1\,pc of their nearest thermally supercritical filament. 
   The mass of each prestellar core appears to be positively correlated with the column density of the local background, itself dominated by the parent filament.
    These findings highlight the tight connection between the cloud's filamentary structure and the distribution of prestellar core masses.
 
   \item  The distribution of measured FWHM widths for the California filaments has a median (undeconvolved) value of $0.18$\,pc 
   using column density data on either side 
   of the filament crests. Given the distance of the California GMC and the resolution of the {\it Herschel} column density data, this is consistent with intrinsic filament widths 
   close to $\sim $0.1\,pc on average, as reported earlier for other nearby clouds. 
   The derived filament column density profiles have a most frequent logarithmic slope of about $-1.1$ and a median logarithmic slope of about $-1.4$, 
   corresponding to logarithmic slopes of $-2.1$ and $-2.4$ for the underlying density profiles, respectively.  
   The median dust temperature along the filament crests is $\sim $15\,K. 
   
   \item The sample of analyzed filaments is estimated to be 80\% complete for line masses above  $\sim 10\,M_\odot\,{\rm pc^{-1}}$. 
   We derived for the first time the distribution of {\it local} masses per unit length for such a complete sample of filamentary structures and 
   found that the corresponding FLMF is consistent with a power-law distribution, $\Delta N/\Delta {\rm log} {M}_{\rm line} \propto {M}_{\rm line}^{-1.5\pm0.2}$, 
   for line masses greater than 10\,$M_\odot \,{\rm pc^{-1}}$. This is 
   similar to the Salpeter power-law IMF, supporting the notion that 
   both the prestellar CMF and the stellar IMF may be partly inherited from the FLMF. 
   
   \item Using an improved toy model for the prestellar CMF in filaments that eliminates assumptions of constant filament length and uniform line mass, 
we established a more direct link between the FLMF and the CMF/IMF. In this toy model, the global prestellar CMF in a molecular cloud corresponds to 
a weighted integration of the CMFs generated by individual filaments within the cloud. 
Overall, our study highlights the role of molecular filaments in shaping the formation and mass distribution of newly born stars. 
\end{enumerate}

\begin{acknowledgements}
This work started in 2018 in the HGBS group of the Astrophysics Department
(DAp/AIM) at CEA Paris-Saclay. After returning to China in 2020, G.-Y.~Zhang 
continued to carry out this work at the China-Argentina Cooperation Station of NAOC/CAS.
G.-Y.~Zhang acknowledges support from a Chinese Government Scholarship (No. 201804910583),  
the China Postdoctoral Science Foundation (No. 2021T140672),
and the National Natural Science foundation of China (No. U2031118). 
This work also received support from the Key Project of International Cooperation of the Ministry of Science and Technology of China under grant number 2010DFA02710, as well as from the National Natural Science Foundation of China under grants 11503035, 11573036, 11373009, 11433008, 11403040, and 11403041.
Ph. A. acknowledges support from CNES, “Ile de France” regional funding (DIM-ACAV+ Program), and 
the French national programs of CNRS/INSU on stellar and ISM physics (PNPS and PCMI). 
We thank the anonymous referee for their positive feedback and insightful comments.
\end{acknowledgements}

\bibliographystyle{aa} 
\bibliography{california.bib} 
\begin{appendix}
\section{Completeness for prestellar cores}
\label{corecomp}

In our investigation of the completeness of prestellar core extraction in the California cloud, we followed a methodology similar to that employed by \citet{Konyves+2015} 
in their study of prestellar cores in Aquila (see Appendix B.2 of their paper). Our approach involved simulating prestellar cores within {\it Herschel} images using radiative transfer models of critical Bonnor-Ebert spheres with well-defined physical properties. These synthetic model cores were incorporated into source-free background images derived from actual observations, effectively generating synthetic {\it Herschel} and column density images that reflected the core characteristics in the California cloud. 

The specific details of our simulation are as follows. We generated synthetic images at {\it Herschel}  wavelengths (70, 160, 250, 350, and 500~$\mu $m) with an angular resolution that matched the observed values of 8.4, 13.5, 18.2, 24.9, and 36.3{\arcsec}. Additionally, we produced a column density map at an 18.2{\arcsec} resolution using the simulated images across the 160 to 500~$\mu $m range. 
In total, we included 1041 synthetic prestellar cores within the simulation. We used a background column density threshold of $3\times 10^{21}\ {\rm cm}^{-2}$ for simulated sources. 
Additionally, column density levels were spaced by a factor of 1.3, with considerations for ratios between model and background column densities to ensure reliable placement of model cores. 
We did not simulate any protostars, focusing solely on prestellar cores. We divided the mass distribution into 25 discrete mass bins. 
The core mass function of the model followed $\Delta N/\Delta {\rm log}M \propto M^{-0.7}$, with a lower mass limit of 0.04\,$M_\odot$ and an upper mass limit of 20\,$M_\odot$.
To optimize the placement of model cores, we sorted them based on their radii. Randomization was employed for cores with masses exceeding a specified limit of 0.5\,$M_\odot$. Additionally, a maximum beam size of 24.9{\arcsec} was used to control any potential overlap between sources.

\begin{figure}[h]
   \centering
   \includegraphics[width=1.0 \hsize]{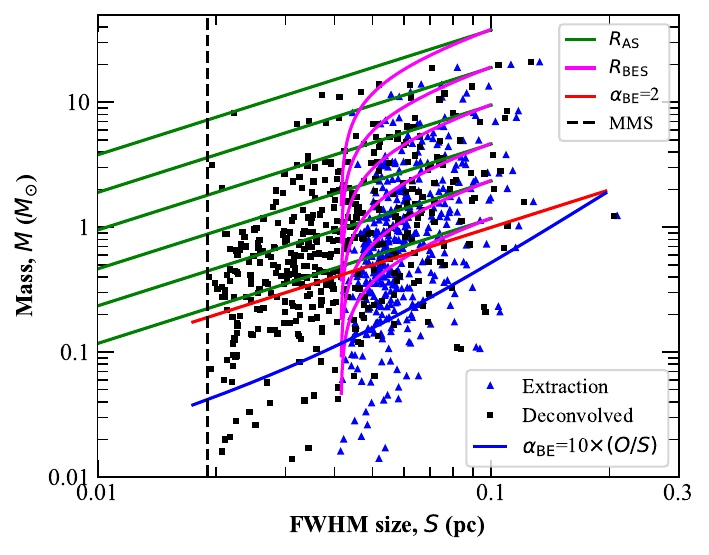}
\caption{Mass--size diagram of the cores detected in the simulated {\it Herschel} images. 
The red line represents half of the critical Bonnor-Ebert mass ($M_{\rm BE}$/2) at 10 K as a function of radius.
The blue line shows the empirical  threshold used to select candidate prestellar cores in Sect.~\ref{selection_core}  (cf. Fig.~\ref{MassvsR}), 
which captures 95\% of the synthetic prestellar cores in the simulation. The curves labeled $R_{\mathrm{BES}}$ (in magenta) show the mass vs. radius relations for the convolved model cores, while the lines labeled $R_{\mathrm{AS}}$ (in green) represent the true mass vs. radius relations of the models. The black dashed vertical line shows the minimum derived model size (MMS) 
used when applying simple Gaussian deconvolution to the simulated extraction results (see Sect.~\ref{selection_core}).}
\label{MassvsRsim}
\end{figure}

We applied the \textsl{getsf} method to extract the model cores within these synthetic images, and we employed the same approach as outlined in Sect.~\ref{selection_core} to select reliable candidate cores. 
As our simulations were conducted using prestellar parameters, the model cores were exclusively self-gravitating prestellar cores. 
Comparison with the extracted sources  shows that
the default selection criterion to select prestellar cores, $\alpha_{\text{BE}} = M_{\text{BE}}^{\text{crit}}/M_{\text{core}} \leq 2$, is overly strict, 
as it only selects $\sim$70\% of the model prestellar cores in our simulated images (as shown in Fig.~\ref{MassvsRsim}). 
In the previous study by \citet{Konyves+2015}, an empirical formula for selecting candidate prestellar cores was introduced: $\alpha _{\text{BE}} \le 5 \times (O/S)^{0.4}$. 
We have adjusted this formula based on our present simulation results to suit the detection outcomes for the California cloud. The modified empirical formula 
is as follows: 
$\alpha _{\text{BE}} \le 10 \times (O/S)$. 
This revised formula, shown as a blue curve in Fig.~\ref{MassvsRsim}, 
successfully identifies 95\% of prestellar cores in our simulation detection results.

\begin{figure*}
   \centering
   \includegraphics[width=0.33 \hsize]{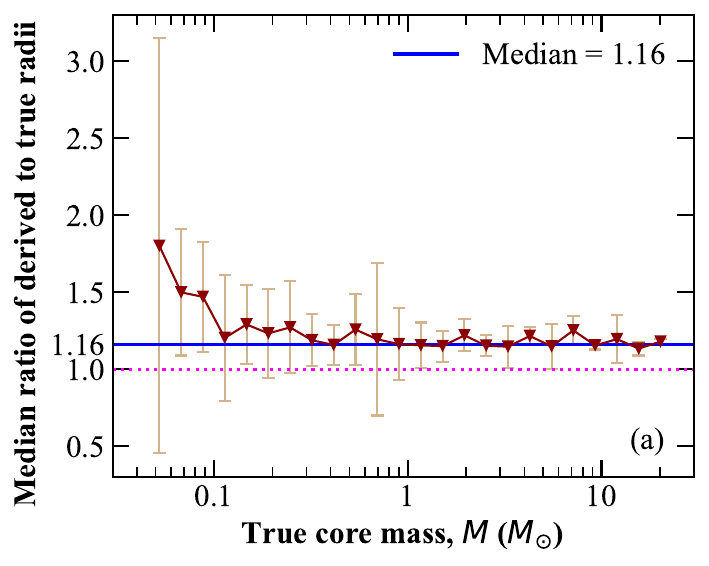}
   \includegraphics[width=0.33 \hsize]{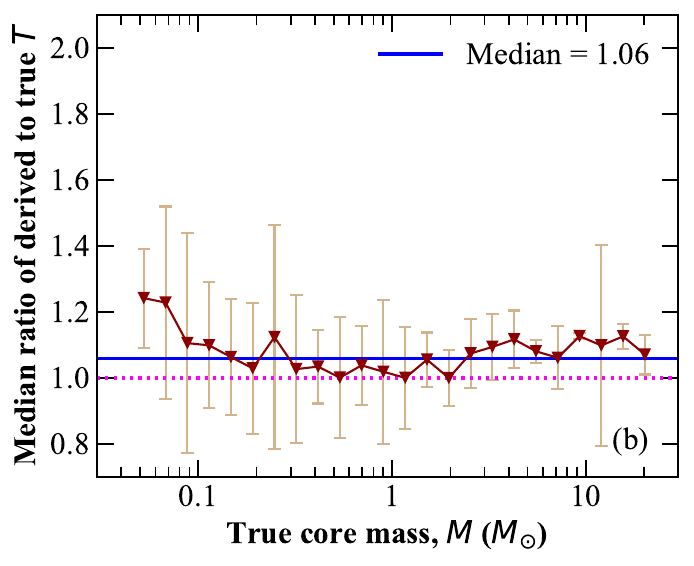}
   \includegraphics[width=0.33 \hsize]{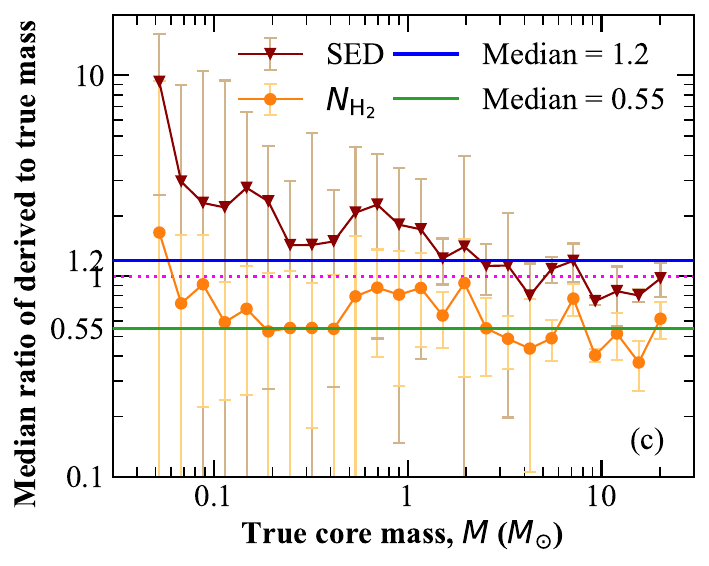}
\caption{Accuracy of core measurements according to the simulations. 
{\bf a)} Measured FWHM radius versus model radius ($R_{\rm BES}$). 
Perfect core radius estimates would lie on the horizontal magenta dotted line. 
The horizontal solid line marks the median ratio value for cores with \( M > 0.25 \, M_{\odot} \).
{\bf b)} SED-derived temperature against true dust temperature of the model.
{\bf c)} SED-derived and $N_{\rm H_{2}}$-based estimates of core mass against true core mass.
}
\label{coreaccuracy}
\end{figure*}

The core masses discussed in the present paper were derived via 
SED fitting of the background-subtracted total fluxes measured 
toward each core. In the literature, another, simpler method has also been used to gauge core masses, by integrating column density over the footprint 
of each core in the $N_{\rm H_{2}}$ map derived from {\it Herschel} data and subtracting the estimated $N_{\rm H_{2}}$ backgound.
For this reason, we tested both approaches using our simulation. 
Figure~\ref{coreaccuracy}\emph{c} clearly shows that SED-derived masses tend to be closer to the true values than $N_{\rm H_{2}}$-based masses. 
As expected, SED-derived masses become more accurate for stronger, more massive cores.
Above $\sim 0.5\, M_\odot$, the median ratio of SED-derived to true core mass is about 20\% on average, indicating a slight overestimation of core masses. 
In comparison, $N_{\rm H_{2}}$-based masses underestimate true masses by approximately 45\% on average.

\begin{figure*}
   \centering
   \includegraphics[width=0.33 \hsize]{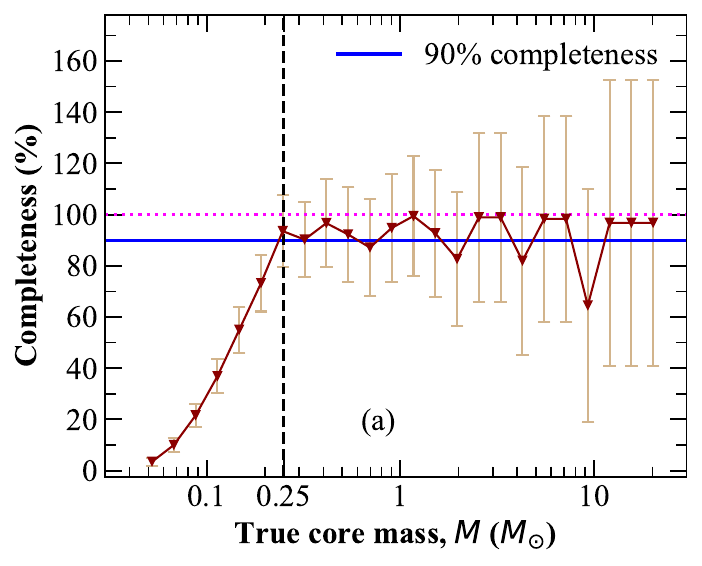}
   \includegraphics[width=0.33 \hsize]{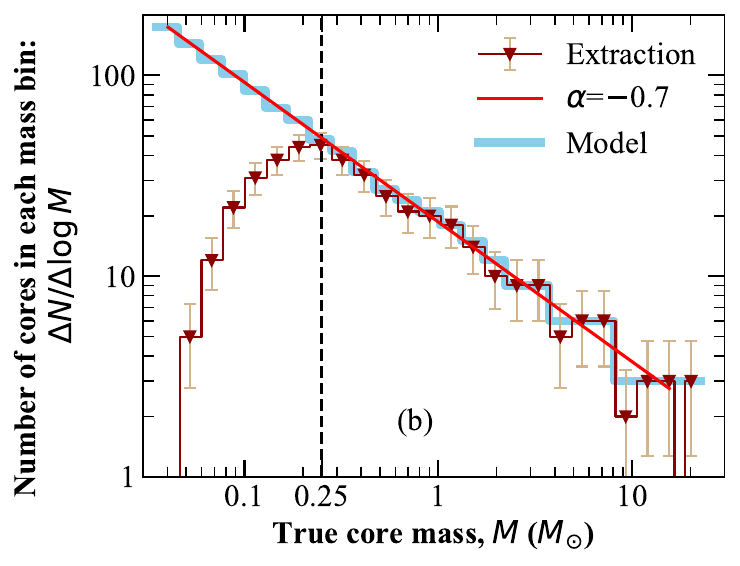}
   \includegraphics[width=0.33 \hsize]{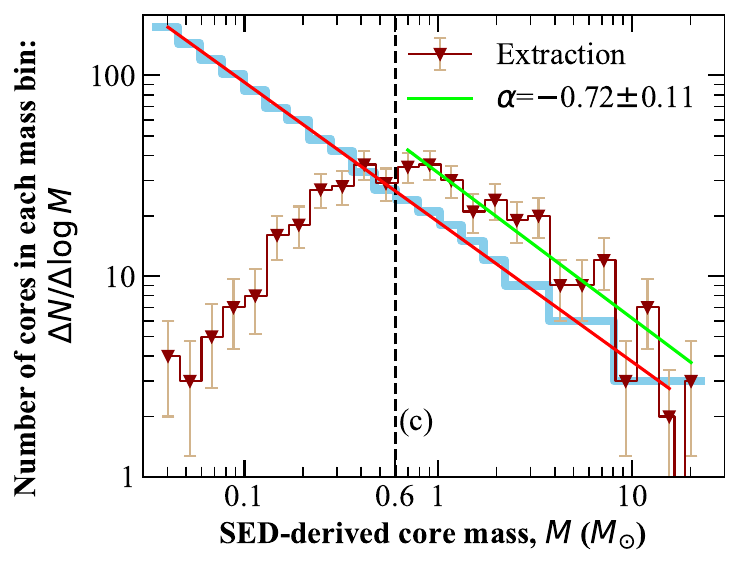}
\caption{Tests for estimating the completeness of core extractions based on the simulations.
{\bf a)} Fraction of recovered synthetic cores as a function of true core mass. The error bars represent the $\pm1\sigma$ deviation within each mass bin and the magenta dotted line marks 100\% completeness.
{\bf b)} Recovered CMF after extracting the simulated cores and binning the results according to true core mass.
{\bf c)} Recovered CMF after extracting the simulated cores and binning the results according to SED-derived mass.
}
\label{corecompleteness}
\end{figure*}

In Fig.~\ref{corecompleteness}\emph{a}, we assess prestellar core completeness by comparing the number of detected cores within each mass bin to the true number of 
model cores. As core mass increases, completeness improves. For true core masses exceeding 0.25\,$M_\odot$, the completeness level exceeds 90\%, fluctuating between 80\% and 100\% thereafter. 
Figure~\ref{corecompleteness}\emph{b} presents the recovered CMF from simulated source extractions, when cores are binned using true masses. 
The recovered CMF matches well the input mass function 
($\Delta N/\Delta {\rm log}M \propto M^{-0.7}$) of the model cores above 0.25\,$M_\odot$,
while the drop below this level is due to significant incompleteness of core detections at lower masses, as shown in Fig.~\ref{corecompleteness}\emph{a}. 
Since the derived core masses are typically overestimated by 20\%, this suggests that the completeness level is $\sim 0.25 \times 1.2 = 0.3\, M_\odot$ in observed core mass.
On the other hand, when cores are binned in terms of SED-derived mass as opposed to true mass, the measured CMF (shown in Fig.~\ref{corecompleteness}\emph{c}) 
is consistent with the input power-law mass function only above $\sim$0.6\,$M_\odot$, suggesting that good completeness is only achieved above 
$\sim $0.6\,$M_\odot$ in observed core mass.
In summary, we conservatively estimate that the actual 90\% completeness level of our {\it Herschel} census of prestellar cores in California  
is in the range of $\sim $0.3--0.6\,$M_\odot$ or $0.45\pm0.15\,M_\odot$ in derived core mass.

\section{Completeness for extracted filaments}
\label{filamentcompleteness}

\begin{figure}
\centering
\includegraphics[width=1.0 \hsize]{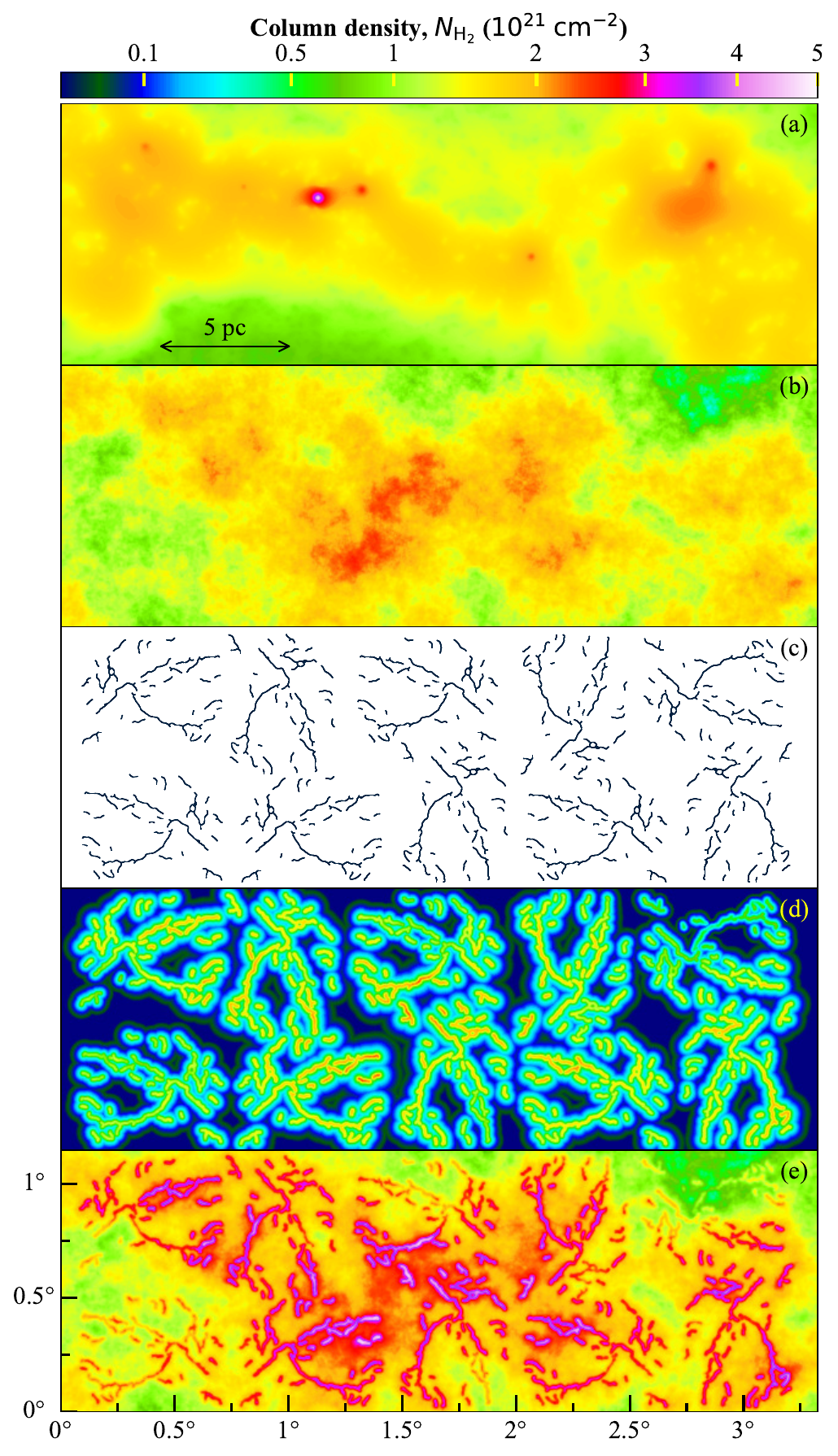}
\caption{Simulated $3.3{\degr}\times 1.1{\degr}$ column density image of a filamentary cloud at $18.2${\arcsec} resolution: (\emph{a}) large-scale background produced by the \textsl{getsf} analysis of the California data; (\emph{b}) synthetic background cloud; (\emph{c}) skeletons detected by \textsl{getsf} in the X-shaped nebula; (\emph{d}) simulated filaments; (\emph{e}) complete synthetic image of the simulated filamentary cloud.}
\label{xshapedsim}
\end{figure}

To assess the completeness of the extracted filamentary structures,  a series of additional simulated images were generated. This process involved the insertion of synthetic filamentary structures with known parameters into a synthetic background image. The synthetic background image, depicted in Fig.~\ref{xshapedsim}\emph{b}, was created by superimposing synthetic cirrus cloud fluctuations spanning all spatial scales onto a smooth background (Fig.~\ref{xshapedsim}\emph{a}) generated by \textsl{getsf}.

To ensure that the simulated filaments closely resembled the lengths and curvatures of the observed filaments, we used the skeletons of the X-shaped nebula region, which were obtained during our filament extraction process in the California cloud (see \citealt{Zhang+2020}). We distributed these skeletons at ten locations within the image area, allowing for rotations of 90{\degr} or 180{\degr} and/or mirroring, as depicted in Fig.~\ref{xshapedsim}\emph{c}. Subsequently, we transformed these skeletons into simulated filaments using the \textsl{modfits} utility, an integral component of the \textsl{getsf} extraction software \citep{Menshchikov2021method}. The transformation was accomplished by applying a prescribed Moffat (Plummer) profile, represented by the equation:
\begin{equation}
N_{{\rm H}_2}(\theta) = N_{0} \left(1 + (2^{-2/s}\!-1) \,\frac{\theta^2}{(W/2)^2}\right)^{s/2},
\label{moffatfun}
\end{equation}
where $\theta$ is the angular distance from the skeleton, $W$ is the filament full width at half maximum, $s < 0$ is the power-law exponent of the column density profile at large radii, 
and $N_0$ is the crest column density. This function has an almost flat, Gaussian inner profile at $\theta < W/2$ that transforms into a power-law profile ${N_{{\rm H}_2}(\theta) \propto \theta^{\,s}}$ at ${\theta \gg W}$. We parameterized the filament strengths by the contrast $C = N_0 / N_{\rm B}$ with respect to the background value. By adjusting the values of $C$ and $W$, we were able to evaluate the ability 
of \textsl{getsf} to detect weaker or wider filaments, thereby allowing us to assess the extraction completeness limits. 

As an illustration, Fig.~\ref{xshapedsim}\emph{d} shows the model filaments (for $C = 1$, $W = 0.15$\,pc, and $s = -1.5$) and Fig.~\ref{xshapedsim}\emph{e} displays the complete simulated image. 
To investigate how the filament detectability depends on contrast and width, we created a set of synthetic images with different values of $C$ (0.1, 0.2, 0.3, 0.4, 0.5, 0.6, 0.8, 1, 2, 4) and $W$ (0.05, 0.08, 0.1, 0.13, 0.15, 0.18, 0.2, 0.3, 0.4\,pc), 
while keeping $s = -1.5$ constant, similar to the average logarithmic slope estimated in the real California data (Fig.~\ref{histogram}). 
The model filaments were extracted independently in each image, adopting the same value for the \textsl{getsf} maximum width parameter as used in the extractions performed
with the observations (cf. Sect.~\ref{sec:filaments}). 
This ensures that the simulations described here accurately reflect our analysis of real data.

\begin{figure}
\centering
\includegraphics[width=1.0 \hsize]{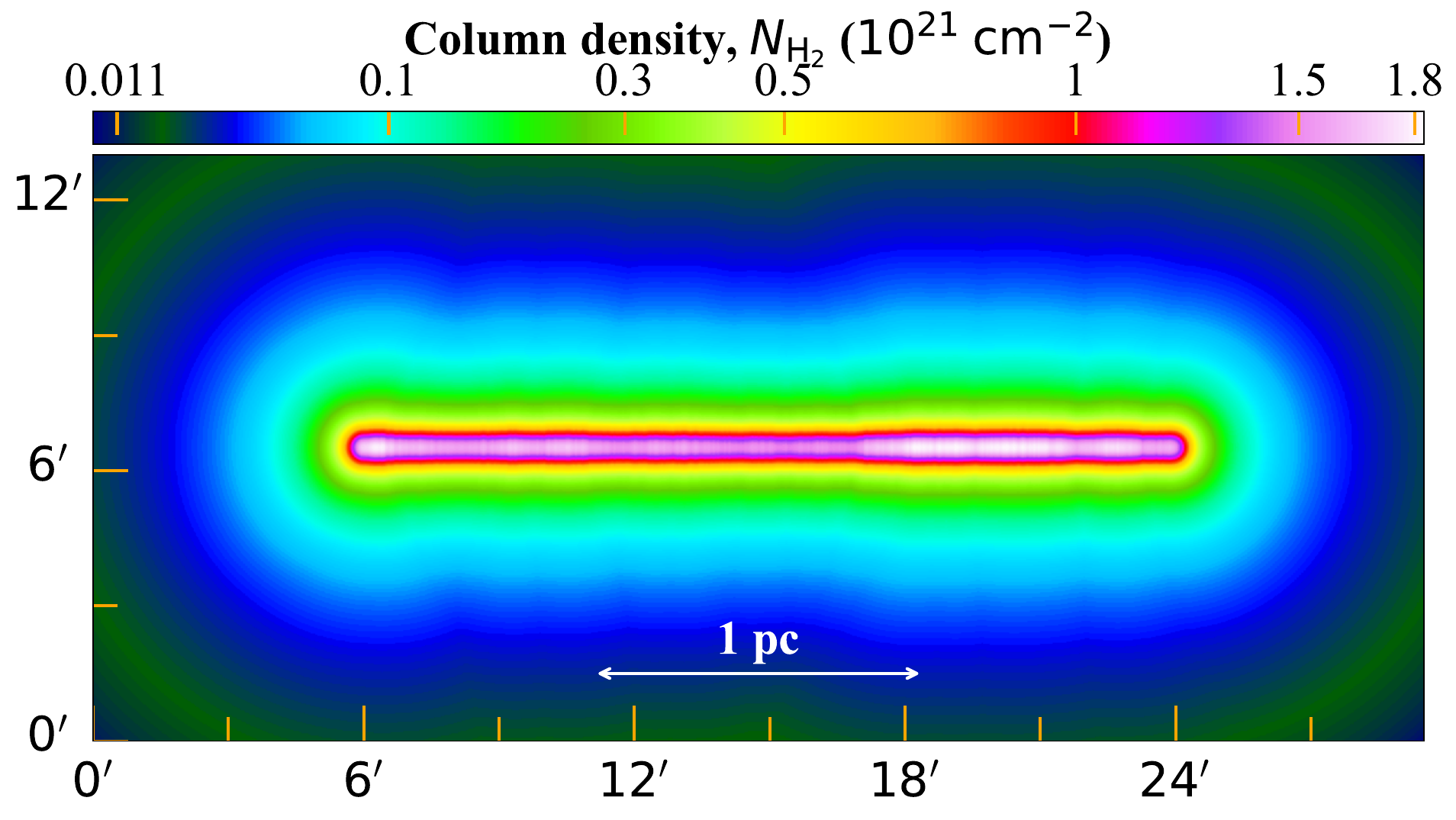}
\caption{Illustrative example of an artificially straightened filament with a width of $W = 0.1$\,pc, a contrast of $C = 1$, and a power-law exponent of $s = -1.5$.}
\label{trueexample}
\end{figure}

For each simulated filament, we used well-defined parameters for the width, contrast, and power law exponent of the filament profile. These fixed parameters represent the true parameters of the simulated filament. 
However, the intrinsic line mass of each simulated filament is not accurately known and needs to be calibrated. 
To do so, we used \textsl{fmeasure} to measure each simulated filament in the absence of background fluctuations and obtain its intrinsic line mass. The filaments display curvature and the images contain more than one filament. To address concerns related to self-blending and overlap with other filaments, 
we artificially straightened the filaments within the images, thereby isolating the filaments from one another. 
An example of such a straightened,  single filament is illustrated in Fig.~\ref{trueexample}. This isolated filament has no background, and its boundaries can extend indefinitely. To mitigate this issue, the \emph{fmeasure} sets an arbitrary boundary where the density is 100 times lower than that of the crest.

\begin{figure}
\centering
\includegraphics[width=1.0 \hsize]{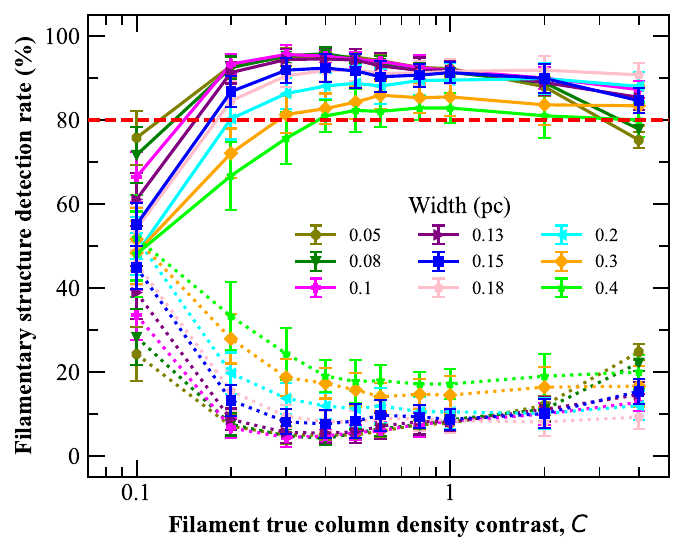}
\caption{Fraction of recovered model filaments with various intrinsic widths (solid lines) 
as a function of column density contrast. 
The red dashed horizontal line marks an 80\% completeness level. 
The dotted lines represent the fraction of spurious filaments.}
\label{detection}
\end{figure}

Analyzing the extraction results, we matched the filament skeletons detected in the simulated images with the true model skeletons. 
The adopted matching radius was $R_{\rm match} = \frac{\sqrt{O^2 + W^2}}{4}$, where $O$ represents the telescope resolution ($\sim$\,0.04\,pc)  and $W$ is true FWHM size of the model filament. The resulting filament detection rates are shown in Fig.~\ref{detection} for a wide range of model filament contrasts. The detection rate is lower for wider filaments, because the wider power-law wings of such filaments imply similarly wider footprints and higher inaccuracies in their background subtraction. 
For model filaments with $W \lesssim 0.15$ pc, $W \lesssim 0.18$ pc, and $W = 0.2$ pc, the detection rate exceeds 90\% at $C \gtrsim 0.2$, $C \gtrsim 0.3$, and $C \gtrsim 0.4$, respectively.
For wider model filaments with $W = 0.3$~pc and $W = 0.4$~pc, the detection rate is higher than 80\% when $C \gtrsim 0.3$ and $C>0.4$, respectively.
Based on the upper envelope of the relation between filament line mass and column density contrast shown in Fig.~\ref{CNM}\emph{b}, 
we therefore estimate that the 80\% detection completeness level of our filament sample is at $\sim $10\,$M_\odot {\rm pc^{-1}}$ 
in terms of derived $M_{\rm line}^{\rm M}$ values.
Since $M_{\rm line}^{\rm M}$ is typically underestimated by $30\pm20$\% when $C>0.4$ (see below), 
this completeness level corresponds to 
$\sim $13\,$M_\odot {\rm pc^{-1}}$ in terms of intrinsic filament line mass. 

\begin{figure}
\centering 
\includegraphics[width=1.0 \hsize]{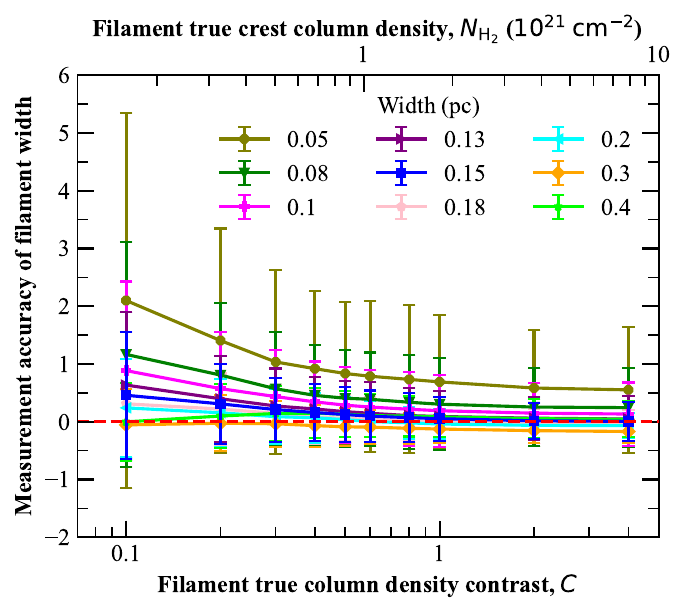}
\caption{Accuracy of measured filament widths as a function of filament crest contrast. The dashed red line marks perfect accuracy.
}
\label{widthaccuracy}
\end{figure}

The simulations described in this Appendix were are also used to assess the accuracy of derived filament properties.
To do so, we used the input model skeletons to measure the filament properties in the filament component map derived from the simulated image, 
and then compared the measurement results with the true properties to evaluate the measurement accuracy of each property.
We define the measurement accuracy $RD$ of a parameter $X$ as the relative difference between the measured value ($X_{\rm m}$) and the true value ($X_{\rm t}$) of the parameter, 
i.e., $RD \equiv \frac{X_{\rm m} - X_{\rm t}}{X_{\rm t}}$. 
Figure~\ref{widthaccuracy} shows the measurement accuracy of filament widths as a function of filament contrast for several intrinsic values of the filament width. 
It can be seen that, at low contrasts $\sim$0.1, 
the filament width estimates can be quite inaccurate. 
However, as the contrast increases, the measurement results progressively converge toward the true widths of the model filaments. 
At filament contrasts higher than 0.4, the average measurement accuracies are  0.73, 0.34, 0.23, 0.12, 0.08, 0.02, $-$0.01, $-$0.12, and 0.1 for filament widths of 0.05, 0.08, 0.1, 0.13, 0.15, 0.18, 0.2, 0.3, and 0.4 pc, respectively.
For model filaments with intrinsic widths of 0.18 and 0.2\,pc, the measurements are very good with errors of less than 2\%. 
For intrinsic widths from 0.13 to 0.4\,pc, the width estimates remain very accurate with errors of less than 10\%. 
For intrinsic widths of 0.08 and 0.1 pc, the measurement errors range from 20 to 35\%. 
For model filaments with intrinsic widths of 0.05\,pc, approaching the telescope resolution of 0.04\,pc, the measurement errors are larger $\sim$\,75\%, 
as expected since it is no longer possible to resolve fine details in the filament structure. 
The measurements of California filaments with intrinsic widths of 0.08 and 0.1\,pc may also be affected by the limited resolution of the {\it Herschel} telescope, 
while wide filaments, such as those with intrinsic widths of 0.3 and 0.4\,pc, might blend together, thereby impacting measurement accuracy.

\begin{figure}
\centering 
\includegraphics[width=1.0 \hsize]{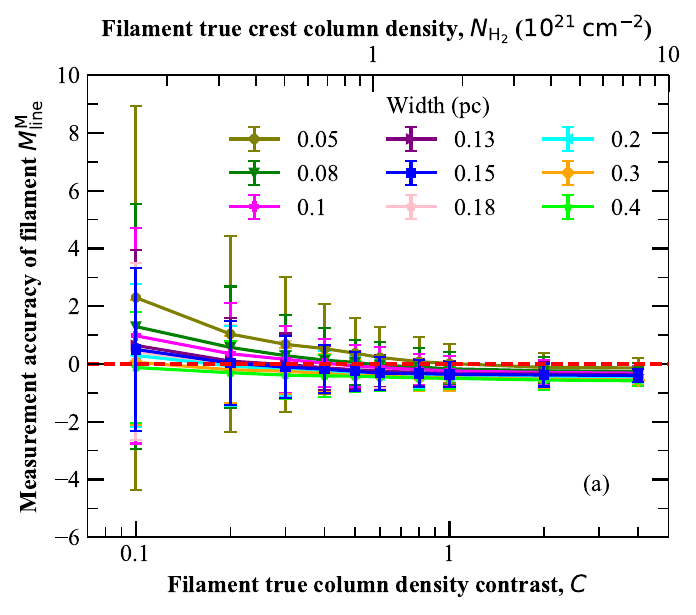}
\includegraphics[width=1.0 \hsize]{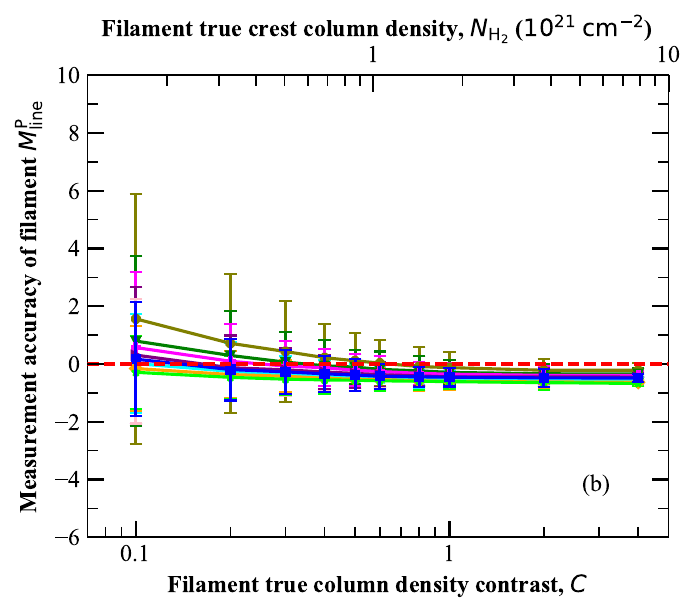}
\caption{Accuracy of ${M}_{\rm line}^{\rm M}$ and ${M}_{\rm line}^{\rm P}$ line mass estimates as a function of filament contrast. The dashed red lines mark perfect accuracy.}
\label{mlineaccuracy}
\end{figure}

Figure~\ref{mlineaccuracy}\emph{a} and \emph{b} displays our estimates of the measurement accuracy for the line mass parameter, $M_{\rm line}^{\rm M}$ or $M_{\rm line}^{\rm P}$. 
At low filament contrast, the measurement error is again relatively large. 
As the contrast increases, both $M_{\rm line}^{\rm M}$ and $M_{\rm line}^{\rm P}$ approach the true line masses. 
When the contrast exceeds 0.4, the average accuracy of $M_{\rm line}^{\rm M}$ estimates of the line mass 
is 0.14, $-0.09$, $-0.17$, $-0.27$, $-0.32$, $-0.37$, $-0.39$, $-0.45$ and $-0.48$,
for model filaments with widths of 0.05, 0.08, 0.1, 0.13, 0.15, 0.18, 0.2, 0.3 and 0.4\,pc  respectively. 
For the same model filaments, the average accuracy of $M_{\rm line}^{\rm P}$ estimates 
is $-0.04$, $-0.22$, $-0.29$, $-0.39$, $-0.43$, $-0.48$, $-0.51$, $-0.57$ and $-0.6$, respectively. 
For filaments contrasts higher than 0.4, it thus appears that $M_{\rm line}^{\rm M}$ is a slightly better estimator of filament line mass than $M_{\rm line}^{\rm P}$. 
However, when the contrast decreases from 0.4 to 0.1, the measurement accuracy of $M_{\rm line}^{\rm M}$ rapidly deterioriates, 
and $M_{\rm line}^{\rm P}$ becomes a slightly better estimator of line mass than $M_{\rm line}^{\rm M}$.
Above a contrast of 0.4, and considering all filament widths, 
$M_{\rm line}^{\rm M}$ typically underestimates the true $M_{\rm line}$ value 
by about $30\pm20$\%, while $M_{\rm line}^{\rm P}$ leads to  $\sim$~$40\pm20$\% underestimates.
\end{appendix}   
\end{document}